\documentclass[iop, onecolumn]{emulateapj}

\slugcomment{Not to appear in Nonlearned J., 45.}

\shorttitle{Pre-supernova neutrino emissions from ONe and Fe cores}
\shortauthors{Kato et al.}

\usepackage{color}
\begin{document}


\title{Pre-supernova neutrino emissions from ONe cores in the progenitors of core-collapse supernovae: are they distinguishable from those of Fe cores?}


\author{Chinami Kato\altaffilmark{1}, Milad Delfan Azari\altaffilmark{1}, Shoichi Yamada\altaffilmark{1,2}, Koh Takahashi\altaffilmark{3}, Hideyuki Umeda\altaffilmark{3}, Takashi Yoshida\altaffilmark{4}, and Koji Ishidoshiro\altaffilmark{5}}
\affil{${}^1$School of Advanced Science and Engineering, Waseda University, \\ 3-4-1, Okubo, Shinjuku, Tokyo 169-8555, Japan}
\affil{${}^2$Advanced Research Institute for Science and Engineering, Waseda University, \\ 3-4-1, Okubo, Shinjuku, Tokyo 169-8555, Japan}
\affil{${}^3$Department of Astronomy, The University of Tokyo, Tokyo 113-0033, Japan}
\affil{${}^4$Yukawa Institute for Theoretical Physics, Kyoto University, Kyoto 606-8502, Japan}
\affil{${}^5$Research Center for Neutrino Science, Tohoku University, Sendai 980-8578, Japan}



\begin{abstract}
Aiming to distinguish two types of progenitors of core collapse supernovae, i.e., one with a core composed mainly of oxygen and neon (abbreviated as ONe core) and 
the other with an iron core (or Fe core), we calculated the luminosities and spectra of neutrinos emitted from these cores prior to gravitational collapse, taking neutrino oscillation into account.  We found that the total energies emitted as $\bar{\nu}_e$ from the ONe core are $\lesssim 10^{46}\ {\rm erg}$, which is much smaller than $\sim 10^{47}\ {\rm erg}$ for Fe cores. 
The average energy, on the other hand, is twice as large for the ONe core as those for the Fe cores. The neutrinos produced by the plasmon 
decays in the ONe core are more numerous than those from the electron-positron annihilation in both cores but they have much lower average energies 
$\lesssim 1\ {\rm MeV}$. Although it is difficult to detect the pre-supernova neutrinos from 
the ONe core even if it is located within 200$\ $pc from the earth, we expect $ \sim 9 - 43$ and $\sim 7 - 61$ events for Fe cores at KamLAND and  Super-Kamiokande, respectively, depending on 
the progenitor mass and neutrino-mass hierarchy. These numbers might be increased by an order of magnitude if we envisage next-generation detectors such as JUNO.  We will hence be able to distinguish the two types of progenitors by the detection or non-detection of the pre-supernova neutrinos if they are close enough ($\lesssim1\ {\rm kpc}$). 
\end{abstract}


\section{Introduction}

Massive stars with the mass $M_{\mathrm{ZAMS}} \gtrsim 8\ \rm{M_\odot}$ on the zero age main sequence are supposed to explode as supernovae at the end of their lives
\citep{janka12, burrows13}. The explosion is actually instigated by implosion of the central core, which is later inverted and leads eventually to the ejection of outer 
envelopes and the formation of compact objects such as neutron stars and black holes. Exactly how that occurs is still a matter of fierce debates (\citealt{janka12, kotake12}
and references therein). The stellar core just prior to collapse will be either consisted of irons (referred to as the Fe core in the following) or composed of oxygen and neon 
mainly (called the ONe core hereafter) \citep{woosley02}. The initial mass $M_{\rm ZAMS}$ is the main factor to determine which is obtained in the end: stars on the 
lightest end of the spectrum of massive stars ($\sim 8 -10\ {\rm M_{\odot}}$) will have the ONe cores in the last stage of their lives whereas more massive stars will produce 
the Fe cores \citep{hashimoto1988,umeda12}. All the Fe cores will eventually collapse gravitationally whereas only a fraction of the ONe cores will implode with the rest resulting in 
white dwarfs \citep{poelarends08,doherty15}. The masses that separate these different regimes are not well determined yet, since these stars commonly experience pulsational 
instabilities that are accompanied by mass ejections toward the end of their lives, which are very difficult to compute numerically \citep{jones13}. 

In the quasi static evolutions of massive stars, the carbon burning produces the ONe cores as an ash. If the star is massive enough ($\gtrsim 10\ {\rm M_{\odot}})$, the central
temperature becomes high enough to ignite these ash elements to produce Si and eventually Fe. In lighter stars, on the other hand, further burnings do not occur because the
ONe core can be supported by degenerate electrons without any energy generation. The core acquires mass thereafter, however, through shell burnings while the outer 
envelopes shedding their masses through the pulsations and winds \citep{jones13}. If the final core mass, which is determined by these competing processes, is small 
and the central density does not reach the threshold $M_{\mathrm core}=1.367\ {\rm M_{\odot}}$, the result will be the formation of the ONe white dwarf as an end point of the 
evolution \citep{koh13}. In the opposite case, i.e., if the core becomes massive enough so that the central density could exceed the threshold of 
$10^{9.88}\ {\rm g/cm^3}$, then electron captures on Mg would commence \citep{koh13}. Once this happens, the core contracts further, 
accompanied by the rise of density and temperature. When the density reaches the critical values of $10^{10.3}\ {\rm g/cm^3}$, the electron captures on Ne, one of the major elements,
start and the implosion is much more accelerated, eventually leading to supernova explosions called the electron-capture supernovae (ECSNe). When the temperature
reaches $10^{9.2}\ {\rm K}$ in the gravitational collapse,  Ne and O are ignited at the center, which is soon followed by Si burnings, and the deflagration starts to propagate 
outward to convert the ONe core to the Fe core. The latter is actually in nuclear statistical equilibrium (NSE) with iron group elements being dominant. The conversion generates
heat and the NSE core becomes much hotter than the ONe core before the conversion.

In the case of more massive stars, the temperatures are higher and electrons are not degenerate until the formation of the Fe core. When the central temperature exceeds 
$10^{9.7}\ {\rm K}$ in the Fe core, photo-dissociations of nuclei take place, which are actually a gradual change in the nuclear abundance in NSE with lighter nuclei becoming 
more abundant. Since this is an endothermic process, the accompanying pressure reduction triggers the collapse of Fe core, which eventually leads to supernova explosions, 
to which we refer as  the Fe-core collapse supernovae or FeCCSNe for short in this paper. 
  
These differences in the stellar evolutions to produce the ONe or Fe core are reflected in the marked differences in the density profiles of the progenitors just prior to the 
core collapse: the ONe cores are in general less massive than the Fe cores; the density decreases very rapidly outside the core and the envelope is much more tenuous
in the ONe cores. As a consequence, the supernova dynamics itself may be different between them. As mentioned already, the mechanism of CCSNe has been a long-standing
unsolved problem for many decades. The initial implosion of the core is reversed by the outward sweep of the shock wave produced by the bounce that occurs for the inner part 
of the core when the central density exceeds the nuclear saturation density ($\sim 3 \times 10^{14}\ {\rm g/cm^3}$). Unfortunately, the shock wave is not strong enough to get 
through the core completely and is stagnated inside the core. Researchers are exploring the way to revive the stalled shock wave and for the moment the so-called neutrino 
heating scenario is supposed to be most promising \citep{janka12}. In this scenario, a fraction of neutrinos emitted copiously from a proto neutron star are re-absorbed by
the matter below the shock wave and the resultant matter-heating reinvigorates the shock wave and produces explosion eventually. The mechanism is notoriously inefficient,
however, and hydrodynamical instabilities of some sorts are believed to be necessary to boost the heating \citep{janka12}. Although there is no consensus in the supernova 
society at present on whether the neutrino heating mechanism really works or not \citep{burrows13}, we are rather certain that it can produce explosion successfully at
least for the ONe core even without the boost by the hydrodynamical instabilities. As a matter of fact, detailed numerical simulations demonstrated \citep{kitaura06,janka07} 
that shock revival occurs rather early thanks to the compact core and the tenuous envelope with the neutrino signals that lack the sign of the instabilities. 
The resultant explosion is believed to be weak with the energy being $\sim 10^{50}\ {\rm erg}$. This may not be a problem for ECSNe, however, since SN1054, which produced 
the Crab pulsar, is supposed to be such an event \citep{nomoto82,tominaga13}.

Neutrinos are crucial not only in the mechanism of CCSNe but also in observations as vindicated in SN1987A \citep{arnet1989}. Indeed CCSNe are one of the most important targets of the
nascent neutrino astrophysics \citep{raffelt12}. It is true that the PNS cooling phase following the shock revival makes the greatest contributions to neutrino emissions. As a matter of
fact, this phase lasts for $\sim 10\ {\rm sec}$ and radiates as neutrinos most of the energy of $\sim 10^{53}\ {\rm erg}$ that are liberated by the gravitational collapse and stored as the internal 
energy in the proto neutron star (PNS) \citep{sato1987,burrows1988,fischer12}. However, neutrino emissions commence much earlier on. It is 
well known in stellar evolution theory that neutrino emissions dominate photon radiations in the stellar cooling after C burning. There are five processes responsible for the 
neutrino emissions: 1. annihilations of electron-positron pairs, 2. plasmon decays, 3. photo-pair processes, 4. bremsstrahlungs by electrons and positrons accelerated by 
nuclei and 5. electron captures on nuclei and free nucleons. Which process is dominant depends on the density and temperature of matter in general \citep{itoh1996} but 
the first two processes are normally the most important in the late phase of the massive star evolution. In particular, the pair annihilation is dominant at high temperatures 
whereas the plasmon decay becomes more important at lower temperatures. 

What is important here is that  the neutrinos emitted by these processes prior to collapse may be observable if the supernova occurs in our vicinity, e.g., within 1kpc. \cite{odrzy04} 
calculated the luminosities and spectra of the neutrinos emitted by the pair annihilations during the C-, Ne-, O-, Si-burnings for a 20 ${\rm M_\odot}$ progenitor model 
with the Monte Carlo method and estimated the detection events for 6 terrestrial neutrino detectors. They found that the mean energies of neutrinos are
$0.71$, $0.97$, $1.1$ and $1.8\ {\rm MeV}$ for the C-, Ne-, O- and Si-burnings, respectively. Assuming that the distance to the supernova is $1\ {\rm kpc}$, they evaluated the
event numbers would be 41 for Super-Kamiokande and 4 for KamLAND.

In this paper, we investigate the possibility to distinguish the ECSN from FeCCSN by the observation of the pre-collapse neutrinos. Employing a latest realistic progenitor model
with an ONe core \citep{koh13} as well as those with Fe cores (private communications with Takahashi), we calculate the temporal evolutions of the luminosities 
and spectra of the neutrinos emitted via the pair annihilations and plasmon decays, the two main emission reactions. We also evaluate roughly the expected numbers of detection 
events both for water Cherenkov detectors and for liquid scintillators, taking neutrino oscillations into account. As the representative detectors of these two types in current operation,
we choose Super-Kamiokande and KamLAND. Although the latter detectors have much smaller fiducial volumes than the former, they have better sensitivities at low energies. 
Note that the average neutrino energies predecited by \cite{odrzy04} are lower than the typical energy thresholds $\gtrsim 5\ {\rm MeV}$ for these large water Cherenkov detectors. 
In addition to these detectors, we also consider Hyper-Kamiokande, a planned next-generation water Cherenkov detector, and JUNO, a larger-scale liquid scintillator under construction. 

The organization of the paper is as follows: the progenitor models for ECSN and FeCCSN are described in Section 2; our method to calculate the luminosities and spectra of
the neutrinos emitted by the pair annihilation and plasmon decay is summarized in Section 3 and detailed in Appendix; the results are presented in Section 4, and finally 
the summary and discussions are given in Section 5.

\section{Progenitor model}

We employed realistic progenitor models with 8.4 \footnote{
This corresponds to $10.8{\rm M_{\odot}}$ model in their original paper by \cite{koh13}. After taking into account convective overshooting in their stellar evolution calculation, they found that the smaller stellar mass produced essentially the same ONe core at the end of the evolution. We have hence decided to adopt this updated number in this paper.
}, $12$ and $15\ {\rm M_{\odot}}$ calculated by Takahashi et al. \citep{koh13}. The first one produces the ONe core that will give 
rise to the ECSN whereas the last two generate Fe cores, which will result in the 
FeCCSNe. The evolution of the $8.4\ {\rm M_{\odot}}$ model is not very different from those of the other two until
the C-burning. Once a core composed of oxygen and neon is formed, they evolve differently. In the lightest model, the central temperature does not reach the value needed to 
ignite Ne, decreasing initially owing to thermally activated neutrino emissions. The ONe core contracts on the Kelvin-Helmholtz time scale as a result. The shell He-burning 
increases the mass of the core, which is now supported by degenerate electrons. As the core becomes fatter, the central density and, as a consequence, the chemical potential 
of electrons also rise there. The $8.4\ {\rm M_{\odot}}$ model is massive enough to exceed the density threshold $\rho = 10^{9.88}\ {\rm g/cm^3}$ for the electron capture on 
$^{24}{\rm Mg}$. The electron capture enhances the contraction of the ONe core. Since the electron captures are exothermic, the temperature increases rather rapidly in this 
phase. When the central density reaches $\rho =10^{10.3}\ {\rm g/cm^3}$, $^{20}{\rm Ne}$ starts to capture electrons, further accelerating the core contraction. When the 
central temperature reaches $10^{9.2}\ {\rm K}$, the O+Ne burning takes place. Then the temperature rises quickly and the nuclear statistical equilibrium, or NSE, is established
at $T \ge 5 \times 10^9\ {\rm K}$. The NSE region expands at $\sim 10^3\ {\rm km/s}$ as the burning of O+Ne propagates outward as a deflagration. Electron captures on iron group 
elements as well as on free protons occur in the NSE region just as in the Fe core of FeCCSNe and finally lead to a rapid collapse of the core.

In the case of more massive progenitors of $12$ and $15\ {\rm M_{\odot}}$, the ONe cores achieve high enough temperatures to ignite Ne and subsequently O and Si stably with
electrons being non-degenerate, to produce Fe cores finally. The masses of these cores 
grow in time as the Si burning continues and the gravitational collapse is induced normally
by photo-dissociations of heavy nuclei at $T \gtrsim 10^{9.7}\ {\rm K}$. It is a rule of thumb that the more massive the progenitor is, the higher temperature they have for a given density.
This is particularly so, however, when comparing $12$ and $15\ {\rm M_{\odot}}$ models with 
the $8.4\ {\rm M_{\odot}}$ model. In the latter model, the core is much cooler until 
the O+Ne deflagration produces NSE, the fact which has consequences in the neutrino emissions in the pre-supernova stages as shown later. The neutrino emissions in the
pre-supernova stages from massive stars that produce Fe cores were studied by some authors \citep{odrzy04}. In this paper we are concerned with the neutrino emissions from the 
ONe cores that will yield ECSNe later. We employed the $12$ and $15\ {\rm M_{\odot}}$ models mainly for comparison to elucidate qualitative differences that the ONe cores
may make.  

\begin{figure}
\plotone{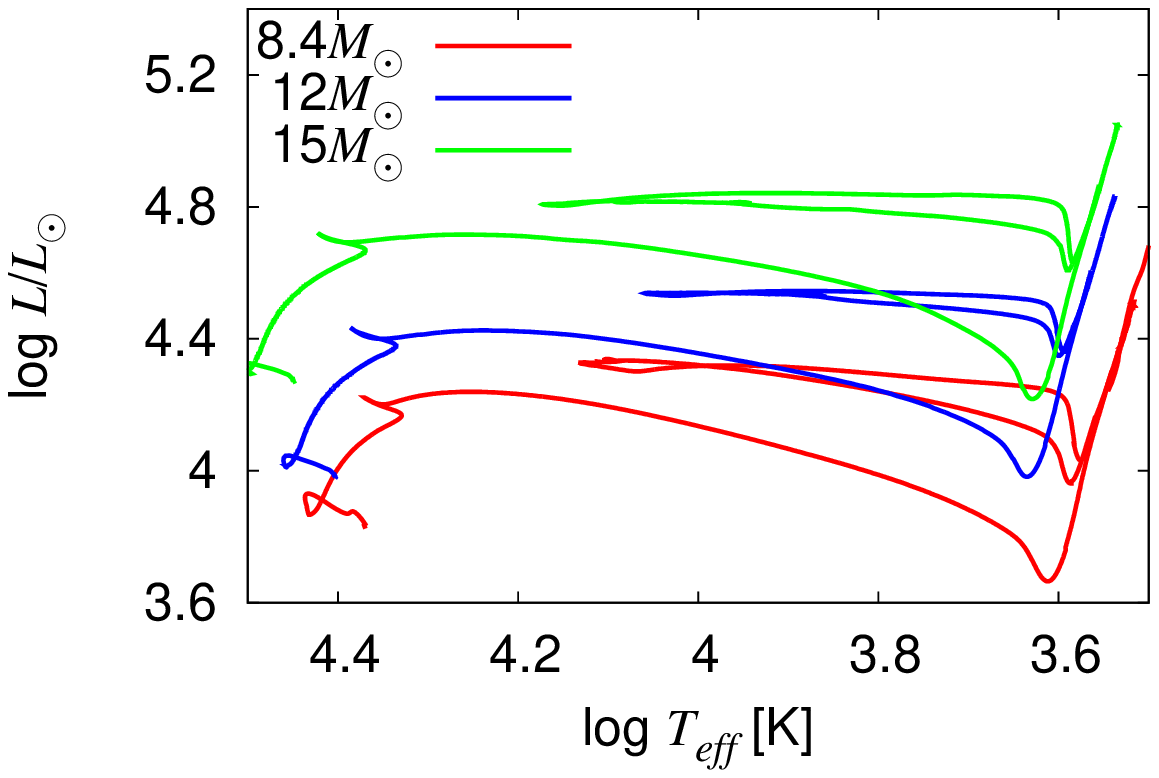}
\caption{The trajectories in the HR-diagram of the progenitors employed in this paper. The red, blue and green curves correspond to 8.4, 12 and $15\ {\rm M_\odot}$ models, respectively.  \label{fig1}}
\end{figure}

\begin{figure}
\plotone{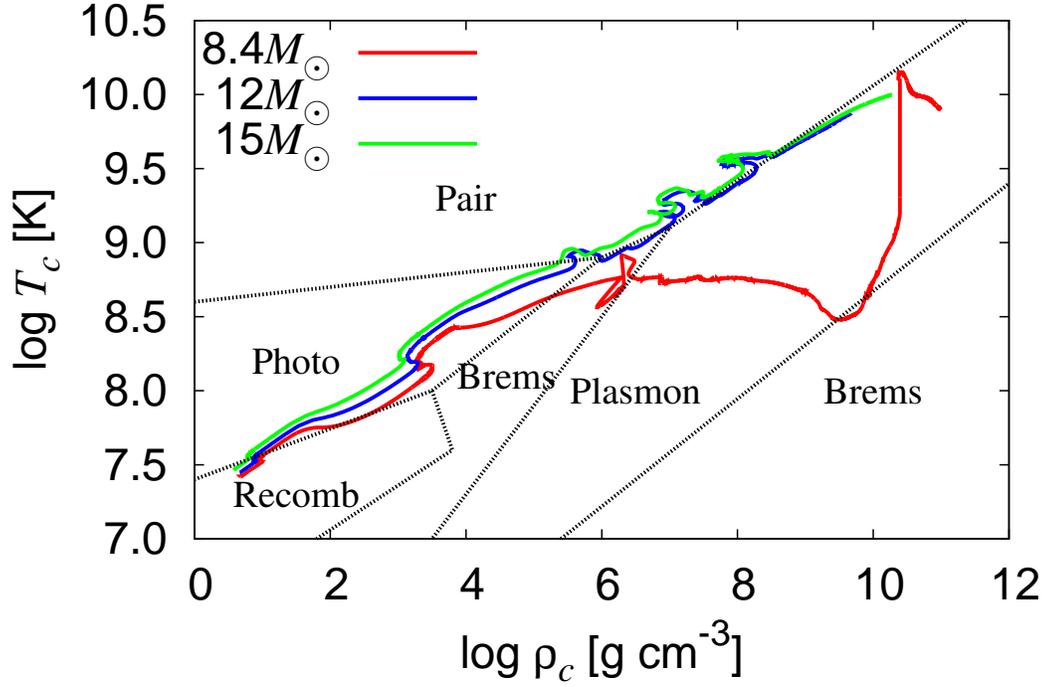}
\caption{The trajectories of the central density $\rho_c$ and temperature $T_c$ for the progenitor employed in this paper. The red, blue and green curves correspond to 8.4, 12 and $15\ {\rm M_\odot}$ models, respectively. Also shown are the boundaries of domains, in which different processes are dominant as indicated \citep{itoh1996}.  \label{fig2}}
\end{figure}

\begin{figure}
\plotone{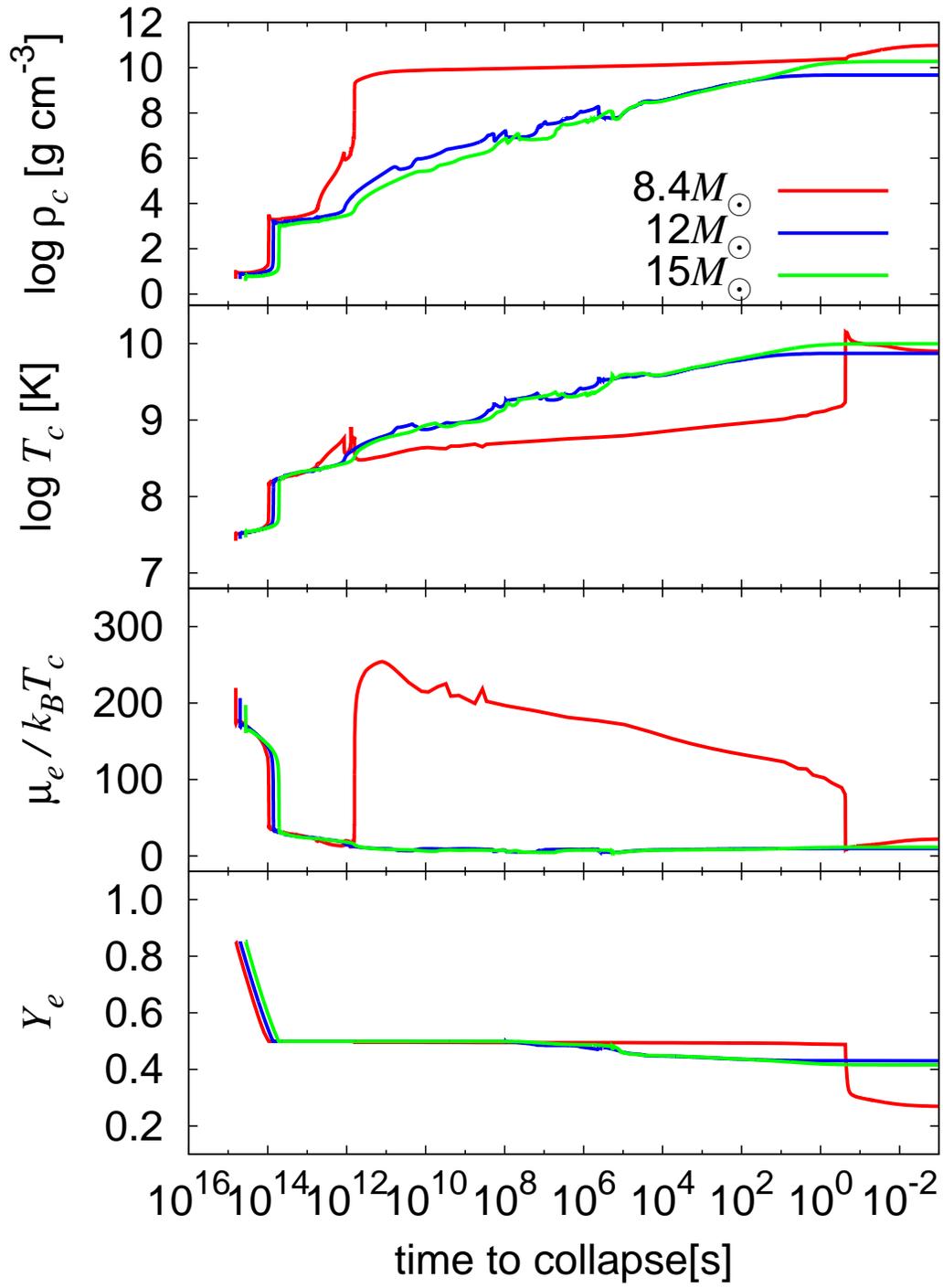}
\caption{The time evolutions of the center central density, temperature, electron degeneracy and electron fraction. The red, blue and green curves correspond to 8.4, 12 and $15\ {\rm M_\odot}$ models, respectively. \label{fig3}}
\end{figure}

Figures~{\ref{fig1}}, {\ref{fig2}} and {\ref{fig3}} present the evolutions of these three progenitor models in different ways. Figure~{\ref{fig1}} is the HR diagram, in which the
temporal changes in the luminosities $L/L_{\odot}$ and effective temperatures $T_{\rm eff}$ are shown as trajectories. It is found that the more massive the star is, the more 
luminous it is but, otherwise, the evolutions are similar to each other. The left panel of 
Fig.~{\ref{fig2}} shows the temporal changes in the central densities and temperatures 
as trajectories in the $\rho - T$ diagram. It is evident that three evolutionary paths are not very different from each other initially up to $\rho_c \sim 10^6\ {\rm g/cm^3}$. As mentioned above, the $8.4\ {\rm M_{\odot}}$ star gets cooled via neutrino emissions efficiently after C-burning and the central temperature becomes much lower than those in other two models, 
for which the central temperatures increase continuously up to the onset of collapse. In the lightest model, the temperature rises rapidly at $\rho_c \sim 10^{10}\ {\rm g/cm^3}$, at which the electron captures first on Mg and then on Ne enhance the core contraction and the O+Ne burning commences at some point, producing the NSE core and 
heating up the matter in it. After NSE is established, the central temperature becomes comparable to or a bit higher than those in the Fe cores of $12$ and $15\ {\rm M_{\odot}}$ models. 
The electron captures on heavy elements accelerate the gravitational collapse that leads to core bounce in the $8.4\ {\rm M_{\odot}}$ model whereas the photo-dissociations of 
iron groups elements trigger the collapse in the other two models. Fig.~{\ref{fig3}} shows the changes in various quantities at the center as functions of time. Note that the horizontal 
axis is the time to collapse. Since it is not easy to define the times of the onset of collapse unambiguously, we choose them simply as the final times in the stellar evolution models, 
at which  $\rho_c = 10^{11.0}\ {\rm g/cm^3}$ and $T_c = 10^{9.9}\ {\rm K}$ for the $8.4\ {\rm M_\odot}$ model whereas $\rho_c = 10^{9.7}\ {\rm g/cm^3}$ and 
$T_c = 10^{9.9}\ {\rm K}$ for the $12\ {\rm M_\odot}$ model and $\rho_c =10^{10.3}\ {\rm g/cm^3}$ and $T_c = 10^{10.0}\ {\rm K}$ for the $15\ {\rm M_\odot}$ model. 
 It is clear again that the $8.4\ {\rm M_{\odot}}$ model is different from the other two qualitatively. This is most manifest in the third panel, in which the 
degeneracy of electrons $\mu_e/k_BT_c$ is shown. Reflecting the fact that the core is much cooler in this model, electrons are strongly degenerate after C-burning. We will find later that this feature 
will cause qualitative differences in the neutrino emissions from these progenitors. It is incidentally pointed out that the electron fraction $Y_e$ drops rather rapidly in the $8.4\ {\rm M_{\odot}}$ 
model once the NSE core is formed by the O+Ne deflagration and electron captures on heavy elements take place.

\begin{figure}
\plotone{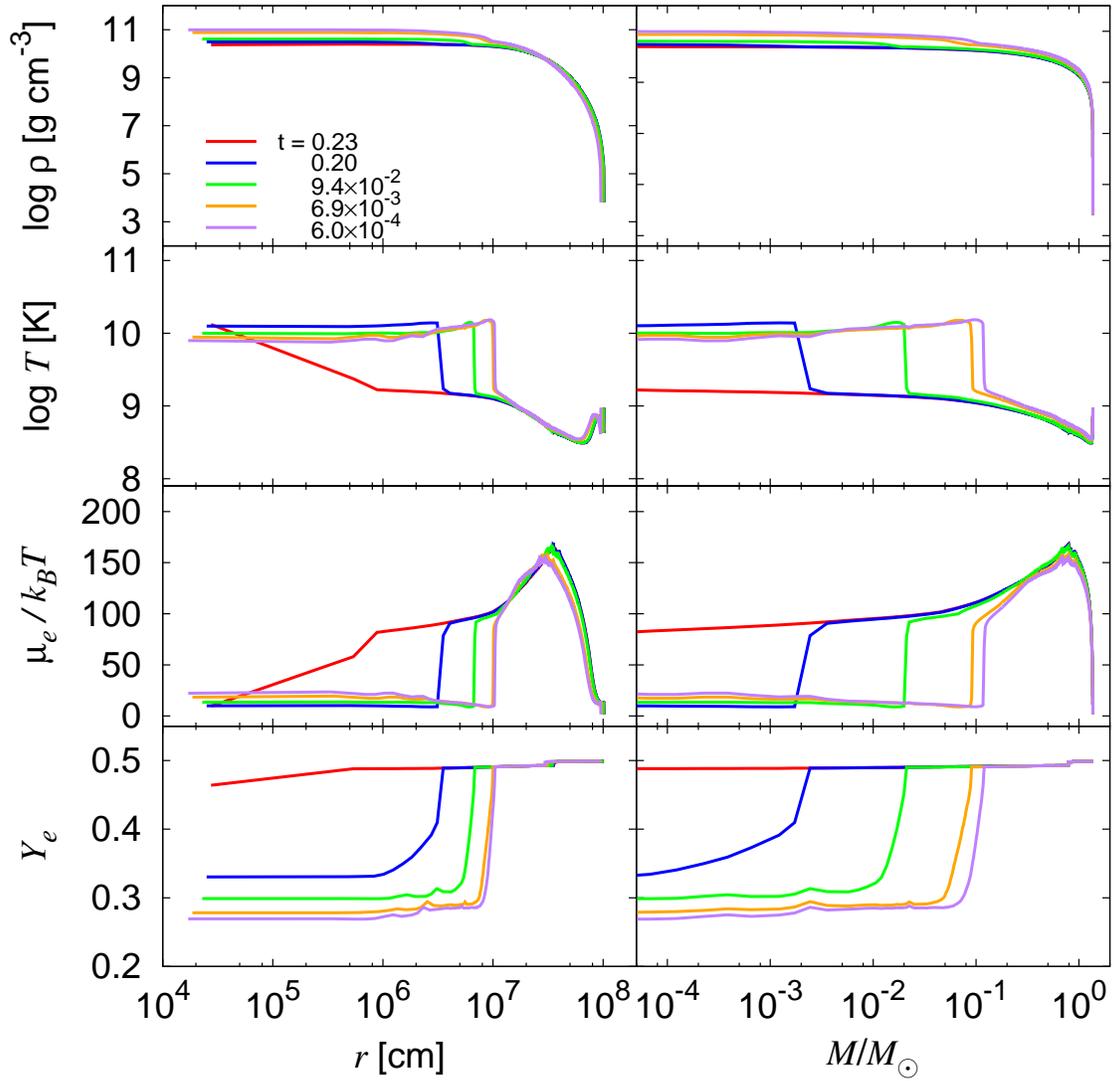}
\caption{The radial profiles of density, temperature, electron degeneracy and electron fraction in the $8.4\ {\rm M_\odot}$ model at different times. Different colors denote different times to collapse:  $t=0.2338\ {\rm s}$ (red line), 0.1972$\ {\rm s}$ (blue line), $9.350\times 10^{-2}\ {\rm s}$ (green line), $6.879\times10^{-3}\ {\rm s}$ (orange line), $5.982\times 10^{-4}\ {\rm s}$ (purple line). \label{fig4}}
\end{figure}

\begin{figure}
\plotone{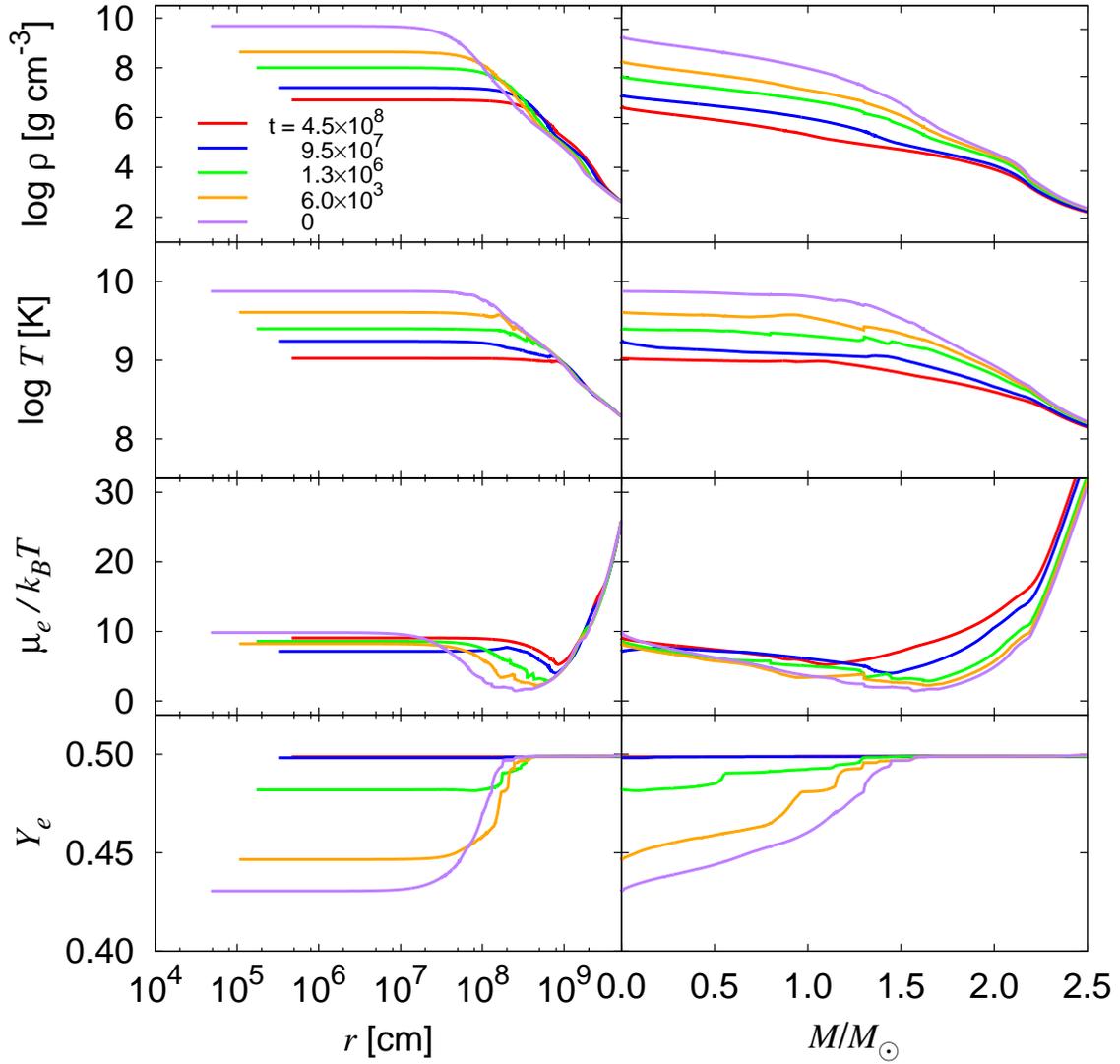}
\caption{The same as Fig. 5 but for the $12\ {\rm M_\odot}$ model. Different colors show different times to collapse: $t=4.537\times 10^8\ {\rm s}$ (red line), $9.490\times10^7\ {\rm s}$ (blue line), $1.344\times 10^6\ {\rm s}$ (green line), $6.044\times10^3\ {\rm s}$ (orange line), 0$\ {\rm s}$ (purple line). \label{fig5}}
\end{figure}

\begin{figure}
\plotone{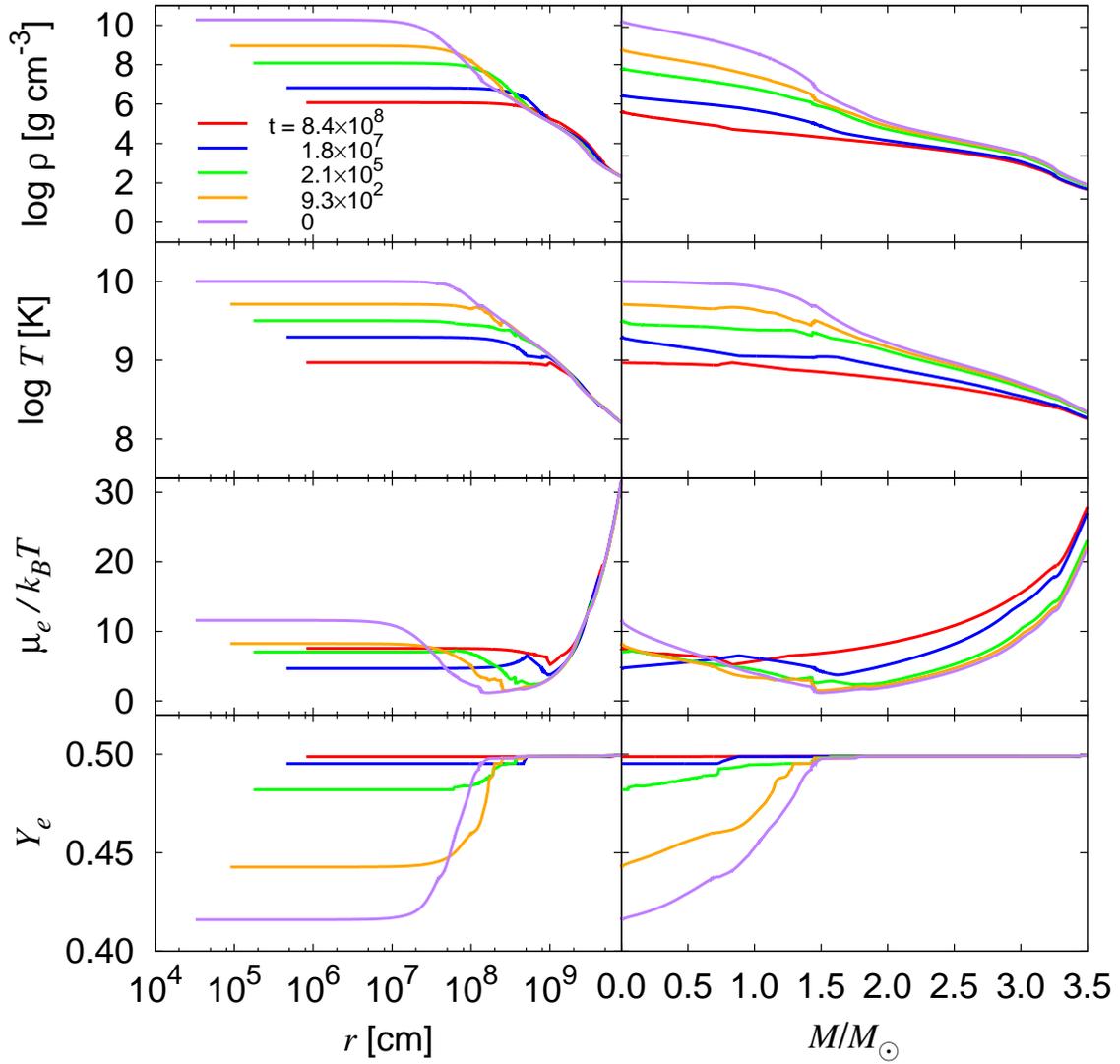}
\caption{The same as Fig. 5 but for the $15\ {\rm M_\odot}$ model. Different colors show different times to collapse: $t=8.431\times 10^8\ {\rm s}$ (red line), $1.824\times10^7\ {\rm s}$ (blue line), $2.099\times 10^5\ {\rm s}$ (green line), $9.340\times10^2\ {\rm s}$ (orange line), 0$\ {\rm s}$ (purple line). \label{fig6}}
\end{figure}

Figures~{\ref{fig4}}, {\ref{fig5}} and {\ref{fig6}} give the radial profiles of density, temperature, electron degeneracy and electron fraction at five different epochs for the three progenitor 
models, respectively. In the left panels these variables are plotted against radius whereas the mass coordinate is used in the right panels. The qualitative differences between the 
$8.4\ {\rm M_{\odot}}$ model and the other two are evident. In fact, we can identify the propagation of the deflagration wave, which sweeps through the ONe core from the center,
producing the hot NSE core behind. It is also recognized that the electron fraction drops only in the NSE region. Figures~{\ref{fig5}} and \ref{fig6} show, on the other hand, that
the electron capture proceeds in the entire Fe cores. In the next section we formulate the neutrino emissivities for the electron-positron annihilation and plasmon decay and evaluate them
for the profiles given here.

\section{Neutrino Emission Rates}

Of the five neutrino-emission processes that are supposed to occur in massive stars, i.e. electron-positron pair annihilation, plasmon decay, photo-pair process, bremsstrahlung 
and electron capture, the first two are more important than the others in general. Having in mind the applications to stellar evolution calculations, \cite{itoh1996} obtained useful fitting 
formulae to the energy loss rates for these processes. They also drew a phase diagram in the
$\rho - T$ plane to indicate which reaction is dominant for a given combination of 
density and temperature. According to their results and the evolutionary paths of our models shown in Fig.~\ref{fig2},
the electron pair annihilation will be dominant for the $12$ and $15\ {\rm M_{\odot}}$ models, which form Fe cores and will hence produce the FeCCSNe whereas the plasmon 
decay will be the most important neutrino-emission process for the $8.4\ {\rm M_{\odot}}$ model, which will result in the ECSN. We will see in the next section that this is indeed 
the case. In order to evaluate the numbers of detection events for the terrestrial neutrino detectors, not only the energy loss rate but also the spectra are needed. In the following,
we describe the formulae that we employed in this work to numerically evaluate the luminosities and spectra of neutrinos for the electron-positron pair annihilation and plasmon decay.

\subsection{electron-positron pair annihilation}

\begin{figure}
\plotone{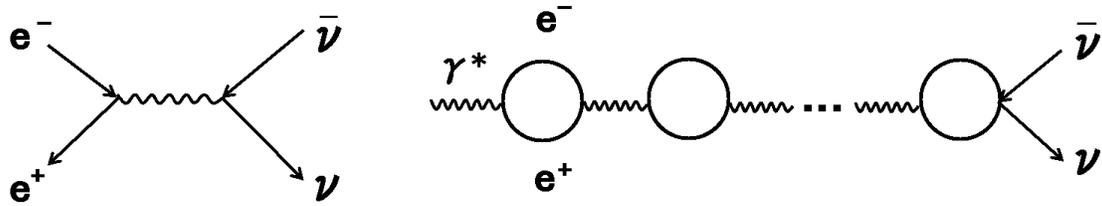}
\caption{Feynman diagrams of the electron-positron pair-annihilation (left panel) and the plasmon decay (right panel). \label{fig0}}
\end{figure}

When the temperature in the core becomes $\gtrsim 10^{9}\ {\rm K}$, the number of photons with high enough energy to produce electron-positron pairs becomes large. As the temperature 
increases, these pairs become highly abundant, being in chemical equilibrium with photons. Although they annihilate each other to generate photons most of the time, they produce 
from time to time pairs of neutrino and anti-neutrino via weak interaction. 
\begin{equation}
\gamma \longleftrightarrow e^+ +e^- \longrightarrow \nu_e +\bar{\nu}_e
\end{equation}
The Feynman diagram corresponding to this process is displayed in Fig.~\ref{fig0}.

The number $R$ of reactions to produce the pair of neutrino and anti-neutrino with four momenta $q^{\mu}=\left(E_\nu, \mbox{\boldmath$q$}\right)$ and $q'^{\mu}=\left(E_{\bar{\nu}}, \mbox{\boldmath$q$}^\prime \right)$, 
respectively, per unit time and unit volume\footnote{$R$ is actually a differential number and $R d^3 q / 2E_\nu d^3 q' / 2E_{\bar{\nu}}$ is the true number of reactions per time 
and volume. See Eq.~(\ref{number}).\label{ft:ft1}} is given by the following equation in the natural unit ($c= \hbar = 1$): 
\begin{equation}
R=\left(\frac{G_{F}}{\sqrt{2}}\right)^2 \! \int \!\!\! \int \!\frac{d^3 k}{\left(2\pi\right)^3 2E_e} \frac{d^3 k^\prime}{\left(2\pi\right)^3 2E_e^\prime} \left(2\pi\right)^4 
\delta^4\left(q+q^\prime-k-k^\prime\right) f_{e^-}\left(E_e\right) f_{e^+}\left(E_e^\prime\right) \, 64 \left|M\right|^2,
\label{eq:eq2}
\end{equation}
in which $G_F=1.166364 \times 10^{-11}\ {\rm MeV^{-2}}$ is the Fermi coupling constant, $k^{\mu}=(E_e, \mbox{\boldmath$k$})$ and $k^{\prime \mu} =(E_e^\prime, \mbox{\boldmath$k$}^\prime)$ are the 4-momenta
for electron and positron, respectively; $f_{e^-}$ and $f_{e^+}$ are the Fermi-Dirac distribution functions of electron and positron, respectively; the matrix element squared for this reaction is
expressed as 
\begin{equation}
\left|M\right|^2 = \left(C_V-C_A\right)^2 \left(q\cdot k\right)\left(q^\prime \cdot k^\prime \right) +\left(C_V+C_A\right)^2 \left(q \cdot k^\prime\right)\left(q^\prime \cdot k\right) +
{m_e}^2\left(C_V-C_A\right)^2\left(q \cdot q^\prime\right)^2.
\end{equation}
In this equation, the coupling constants are given as $C_V = 1/2 + 2 \sin^2{\theta_w}$ and $C_A = 1/2$ with $\sin^2{\theta_w} = 0.2224$ for the Weinberg angle $\theta_w$. Note that
all neutrinos are assumed to be massless, which is well justified for our purposes.

The expression of $R$ in Eq.~(\ref{eq:eq2}) can be cast into the following form:
\begin{equation}
R=\displaystyle{\frac{8G_F^2}{\left(2\pi\right)^2} \left[\, \beta_1\, I_1 + \beta_2\, I_2 + \beta_3\, I_3\, \right]}
\end{equation}
In the above expression, $\beta$'s are the following combinations of the coupling constants: $\beta_1=\left(C_V-C_A\right)^2$, $\beta_2=\left(C_V+C_A\right)^2$ and $\beta_3=C_V^2-C_A^2$, 
and $I$'s are the functions of the energies of emitted neutrino $E_{\nu}$ and anti-neutrino $E_{\bar{\nu}}$ and the angle $\theta$ between their momenta $\mbox{\boldmath$q$}$ and $\mbox{\boldmath$q$}^\prime$: 
\begin{eqnarray}
&&I_1\left(E_{\nu}, E_{\bar{\nu}}, \cos{\theta} \right) = -\frac{2\pi T \, E_{\nu}^2 {E_{\bar{\nu}}}^2 \left(1-\cos{\theta}\right)^2}{\left[\exp (E_{\nu}+E_{\bar{\nu}}) / T - 1\right]{\Delta_e}^5}  
\left \{ AT^2 \left( \left[G_2\left(y_{\mathrm{max}}\right) - G_2\left(y_{\mathrm{min}}\right) \right] \right. \right. \nonumber \\
&&\ \ \ \ \ \ \ \ \ \ \ \ \ \ \ \ \ \ \ \ \ \  \left. \left. + \left[2y_{\mathrm{max}}G_1\left(y_{\mathrm{max}}\right) - 2y_{\mathrm{min}}G_1\left(y_{\mathrm{min}}\right)\right] + 
\left[y_{\mathrm{max}}^2G_0\left(y_{\mathrm{max}}\right) - y_{\mathrm{min}}^2G_0 \left(y_{\mathrm{min}} \right) \right] \right)  \right. \nonumber \\
&& \ \ \ \ \ \ \ \ \ \ \ \ \ \ \ \ \ \ \ \ \ \ \   + BT \left(\left[G_1\left(y_{\mathrm{max}}\right) - G_1\left(y_{\mathrm{min}}\right)\right] + \left[y_{\mathrm{max}}G_0\left(y_{\mathrm{max}}\right)
- y_{\mathrm{min}}G_0\left(y_{\mathrm{min}} \right) \right] \right) \nonumber \\
&& \ \ \ \ \ \ \ \ \ \ \ \ \ \ \ \ \ \ \ \ \ \ \  \left. +C\left[G_0\left(y_{\mathrm{max}}\right)-G_0\left(y_{\mathrm{min}}\right)\right] \right \}, \\
&&I_2=I_1\left(E_{\bar{\nu}},E_{\nu},\cos{\theta}\right), \\
&&I_3=-\frac{2\pi  T \, m_e^2\, E_{\nu}{E_{\bar{\nu}}}\left(1-\cos{\theta}\right)}{\left[\exp (E_{\nu}+E_{\bar{\nu}})/T - 1\right] \Delta_e}
\left[G_0\left(y_{\mathrm{max}}\right)-G_0\left(y_{\mathrm{min}}\right)\right], 
\end{eqnarray}
with 
\begin{eqnarray}
{\Delta_e}^2& \equiv & {E_{\bar{\nu}}}^2+{E_{\nu}}^2+2E_{\nu}E_{\bar{\nu}}\cos{\theta}, \\
A & = & {E_{\bar{\nu}}}^2+{E_{\nu}}^2-E_{\nu}E_{\bar{\nu}}\left(3+\cos{\theta}\right),  \\
B & = & -2{E_{\nu}}^2+{E_{\bar{\nu}}}^2\left(1+\cos{\theta}\right)+E_{\nu}E_{\bar{\nu}}\left(3-\cos{\theta}\right), \\
C & = & \displaystyle{\left(E_{\nu}+E_{\bar{\nu}}\cos{\theta}\right)^2-\frac{1}{2}{E_{\bar{\nu}}}^2\left(1-\cos{\theta}^2\right)
- \frac{1}{2}\left(\frac{m_e\Delta_e}{E_{\nu}}\right)^2\frac{1+\cos{\theta}}{1-\cos{\theta}} },
\end{eqnarray}
and $\eta^{\prime}=\left(\mu_e+E_{\nu}+E_{\bar{\nu}}\right)/T$, $\eta=\mu_e/T,y_{\mathrm{max}}=E_{\mathrm{max}}/T$, $y_{\mathrm{min}}=E_{\mathrm{min}}/T$ and 
$G_n\left(y\right)\equiv F_n\left(\eta^{\prime}-y\right)-F_n\left(\eta-y\right)$, in which the Fermi integral $F_n(z)$ is defined as 
\begin{equation}
F_n\left(z \right)=\int_{0}^{\infty}\frac{x^n}{e^{x-z}+1}dx;
\end{equation} 
$\eta_e$ denotes the chemical potential of electron; the Boltzmann's constant is taken to be unity in these and following equations. The detail of the derivation of these 
expressions is given in Appendix.

The number spectrum for the neutrino or anti-neutrino (denoted by $\nu_1$) is expressed as an integral of $R$ over the momentum of the partner (referred to as $\nu_2$) as follows:
\begin{equation}
\frac{dQ_N^{\nu_1}}{dE_{\nu_1}} = \frac{E_{\nu_1}}{\left(2\pi \right)^2} \int \!\!\! \int  \frac{d^3 q_{\nu_2}}{\left(2 \pi \right)^3 2E_{\nu_2}} R(E_\nu, E_{\bar{\nu}}, \cos \theta). \label{spectrum}\\
\label{eq:eq-dLN}
\end{equation} 
Note that $\nu_1$ may be $\nu$ or $\bar{\nu}$  and the natural unit is employed here. The corresponding energy spectrum is just obtained as 
\begin{equation}
\frac{dQ_E^{\nu_1}}{dE_{\nu_1}} = E_{\nu_1} \frac{dQ_N^{\nu_1}}{dE_{\nu_1}}.
\end{equation} 
The total number emissivity is found by further integrating $dQ_N^{\nu_1}/dE_{\nu_1}$ over $E_{\nu_1}$ as  
\begin{equation}
Q_N^{\nu_1}= \int dE_{\nu_1} \frac{dQ_N^{\nu_1}}{dE_{\nu_1}} = \int \!\!\! \int \! \frac{d^3 q}{\left(2 \pi\right)^3 2E_\nu} \frac{d^3q'}{\left(2\pi \right)^3 2E_{\bar{\nu}}} R(E_\nu, E_{\bar{\nu}}, \cos \theta), 
 \label{number}
\end{equation}
and the corresponding energy emssivity $Q_E$ is obtained similarly:
\begin{equation}
Q_E^{\nu_1}= \int dE_{\nu_1} \frac{dQ_E^{\nu_1}}{dE_{\nu_1}} = \int \!\!\! \int \! \frac{d^3 q}{\left(2 \pi\right)^3 2E_\nu} \frac{d^3q'}{\left(2\pi \right)^3 2E_{\bar{\nu}}} E_{\nu_1} R(E_\nu, E_{\bar{\nu}}, \cos \theta)  \label{LE}.
\end{equation}
Finally the energy loss rate for this reaction is given by the sum of the energy emissivities over all neutrino species: 
\begin{equation}
Q = \sum_{\nu}^{\rm all species} Q_E^{\nu} \label{energy}.
\end{equation}

The left panels of Fig.~\ref{fig7} show the number spectra, Eq.~(\ref{eq:eq-dLN}), of electron-type anti-neutrino for different combinations of $\rho Y_e$ and $T$. In the top panel, the value of 
$\rho Y_e$ is varied around $\rho Y_e = 10^{10}\ {\rm g/cm^3}$ with the value of $T$ being fixed to $T=10^{10}\ {\rm K}$, whereas the latter is changed for a fixed value of the former 
($\rho Y_e = 10^{10}\ {\rm g/cm^3}$) in the bottom panel. It is evident that the emissivity is very sensitive to temperature. In fact, as the temperature increases by $\sim 20\ {\rm \%}$, the peak number 
luminosity becomes greater by an order of magnitude. This is simply because the number of electron-positron pairs increases as $\propto T^{3}$. It is also observed that the average energy increases
as the temperature rises. The dependence on $\rho Y_e$ is much less drastic: the emissivity decreases with the value of $\rho Y_e$, since the number of electron-positron pairs is reduced in this
case. 

For a later comparison, we give here the fitting formula for the energy loss rate $Q$ proposed by \citep{itoh1996}:  
\begin{eqnarray}
Q_{\rm pair} & = & \frac{1}{2}\left[\left({C_V}^2 + {C_A}^2\right) + 2\left({C^\prime_V}^2+{C^\prime_A}^2\right)
+ \left\{\left({C_V}^2 - {C_A}^2\right) + 2\left({C^\prime_V}^2-{C^\prime_A}^2\right)\right\} q_{\mathrm{pair}}\right] \nonumber \\  
& \times & g(\lambda) \, e^{-\frac{2}{\lambda}} \, f_{\mathrm{pair}}, 
\label{itoh}
\end{eqnarray}
In which the coupling constants are defined as $C^{\prime}_V = 1 - C_{V}$ and $C^{\prime}_A = 1 - C_{A}$, 
and $q_{\mathrm{pair}}$, $g\left(\lambda\right)$ and $f_{\mathrm{pair}}$ are expressed, respectively, as
\begin{eqnarray}
q_{\mathrm{pair}} & = & \left(10.7480\lambda^2 + 0.3967 \lambda^{0.5} + 1.0050\right)^{-1.0} \nonumber \\
&&\times \left[1+\left(\frac{\rho}{\mu_e} \right)\left(7.692\times 10^7\lambda^3 + 9.715 \times 10^6 \lambda^{0.5}\right)^{-1.0}\right]^{-0.3}, \\
g\left(\lambda\right) & = & 1-13.04\lambda^2+133.5\lambda^4+1534\lambda^6+918.6\lambda^8, \\
f_{\mathrm{pair}} & = & \frac{\left(a_0+a_1\xi+a_2\xi^2\right)e^{-c\xi}}{\xi^3+b_1\lambda^{-1}+b_2\lambda^{-2}+b_3\lambda^{-3}},
\end{eqnarray}
with 
\begin{eqnarray}
\lambda & = & \frac{T}{5.9302\times10^9\ \mathrm{K}}, \\
\xi & = & \left(\frac{\rho/\mu_e}{10^9\mathrm\ {g/cm}^{3}}\right)^{\frac{1}{3}}\lambda^{-1}.
\end{eqnarray}
This fitting formula is supposed to be accurate within {10\ \%} of error in the regime, where the electron-positron pair annihilation is dominant over other neutrino-emitting processes. In the left panel of 
Fig.~\ref{fig8} we compare the energy loss rates obtained by the formulae given above (Eqs.~(\ref{LE}) and (\ref{energy})) with those given by the fitting formula, (Eq.~(\ref{itoh})) for different densities 
and a fixed temperature ($T=10^{10}\ {\rm K}$) and electron fraction ($Y_e=0.5$). Only the contribution from electron-type neutrinos is taken into account in this comparison. It is apparent that 
they are in excellent agreement except at high densities $\rho \gtrsim 10^{10}\ {\rm g/cm^3}$, where the electron-positron annihilation is no longer dominant and the fitting formula is not accurate.

\subsection{plasmon decay}
Plasmons are quantized collective motions of plasma. They are much like photons as shown in the Feynman diagram given in the right panel of Fig.~\ref{fig0}, obeying the Bose-Einstein 
statistics. Unlike ordinary photons in vacuum, however, they are massive quasi-particles, having a longitudinal mode in addition to the two transverse modes\footnote{In some textbooks,
only the longitudinal mode is called plasmon. In this paper, we refer also to the transverse modes as plasmon.}. Thanks to this property, a plasmon decay to two massless particles is not
kinetically forbidden, the fact which is in sharp contrast to photons in vacuum. The plasmon decay to a pair of neutrinos, 
\begin{equation}
\gamma^* \longrightarrow \nu + \bar{\nu},
\end{equation}
is one of the main cooling processes in massive star after C-burning. As shown later, it is the dominant neutrino-emitting reactions in the ONeMg core until NSE is established. 

We calculate the neutrino emissivity via the plasmon decay, following \cite{braaten93}. The detailed derivations are given in Appendix. The number of reactions, R, to produce the pair
of neutrino and anti-neutrino with the energies $E_{\nu}$ and $E_{\bar{\nu}}$, respectively, per unit time and volume, $R$, is given as\footnote{See footnote~{\ref{ft:ft1}}.} 
\begin{eqnarray}
&&R=\left(\frac{G}{\sqrt{2}}\right)^2\frac{16{C_V}^2}{e^2}\frac{2{E_\nu}^2{E_{\bar{\nu}}}^2\left(1-\cos{\theta}\right)}{\left[1-\exp\left(E_\nu+E_{\bar{\nu}}\right)T\right]}
\nonumber \\
&&\times \left \{\frac{3{\omega_p}^2}{{\Delta_e}^2} \delta\left(f_L\left(E_\nu, E_{\bar{\nu}}, \cos{\theta}\right)\right)
\left[\frac{E_\nu+E_{\bar{\nu}}}{2\Delta_e}\ln{\frac{E_\nu+E_{\bar{\nu}}-\Delta_e}{E_\nu+E_{\bar{\nu}}+\Delta_e}} +1\right]  \right. \nonumber \\
&&\left. \ \ \ \ \times \left[-2\cos{\theta}\left(E_\nu+E_{\bar{\nu}}\right)^2-2E_\nu E_{\bar{\nu}}{\sin{\theta}}^2+\frac{2\left(E_\nu+E_{\bar{\nu}}\right)^2}{{\Delta_e}^2
}\left(E_\nu+E_{\bar{\nu}}\cos{\theta}\right)\left(E_{\bar{\nu}}+E_\nu\cos{\theta}\right)\right]  \right. \nonumber \\
&&\left. \ \ -\frac{3{\omega_p}^2\left(E_\nu+E_{\bar{\nu}}\right)^2}{{\Delta_e}^2} \delta\left(f_T\left(E_\nu, E_{\bar{\nu}}, \cos{\theta}\right)\right)
\left[1+\frac{E_\nu E_{\bar{\nu}}\left(1-\cos{\theta}\right)}{\left(E_\nu+E_{\bar{\nu}}\right)\Delta_e}\ln{\frac{E_\nu+E_{\bar{\nu}}-\Delta_e}{E_\nu+E_{\bar{\nu}}+\Delta_e}}\right] \right. \nonumber \\
&&\ \ \ \ \left.  \times\left[1-\frac{\left(E_\nu\cos{\theta}+E_{\bar{\nu}}\right)\left(E_{\bar{\nu}}\cos{\theta}+E_\nu\right)}{{\Delta_e}^2}\right]  \right \}
\end{eqnarray}
with the following $f_L\left(E_\nu, E_{\bar{\nu}}, \cos{\theta}\right)$ and $f_T\left(E_\nu, E_{\bar{\nu}}, \cos{\theta}\right)$:
\begin{eqnarray}
&&f_L\left(E_\nu, E_{\bar{\nu}}, \cos{\theta}\right)=2E_\nu E_{\bar{\nu}}\left(1-\cos{\theta}\right) \nonumber \\
&&\ \ \ \ \ \ \ \ \ \ \ \ \ \ \ \ \ \ \ \ \ \ \ +3{\omega_p}^2\frac{2E_\nu E_{\bar{\nu}}\left(1-\cos{\theta}\right)}{{\Delta_e}^2}
\left \{\frac{E_\nu+E_{\bar{\nu}}}{2\Delta_\nu}\ln{\frac{E_\nu+E_{\bar{\nu}}-\Delta_e}{E_\nu+E_{\bar{\nu}}+\Delta_e}}+1 \right \} \\
&&f_T\left(E_\nu, E_{\bar{\nu}}, \cos{\theta}\right)=2E_\nu E_{\bar{\nu}}\left(1-\cos{\theta}\right) \nonumber \\
&&\ \ \ \ \ \ \ \ \ \ \ \ \ \ \ \ \ \ \ \ \ \ \ -\frac{3}{2}{\omega_p}^2\frac{\left(E_\nu+E_{\bar{\nu}}\right)^2}{{\Delta_e}^2}
\left[1+\frac{E_\nu E_{\bar{\nu}}\left(1-\cos{\theta}\right)}{\left(E_\nu+E_{\bar{\nu}}\right)\Delta_e}\ln{\frac{E_\nu+E_{\bar{\nu}}-\Delta_e}{E_\nu+E_{\bar{\nu}}+\Delta_e}}\right]
\end{eqnarray}
Note that the dispersion relations of the longitudinal and transverse modes are obtained from $f_L\left(E_\nu, E_{\bar{\nu}}, \cos{\theta}\right)=0$ and $f_T\left(E_\nu, E_{\bar{\nu}}, \cos{\theta}\right)=0$,
respectively.

The number spectrum $dQ^\nu_N/dE_\nu$ and total emissivities $Q^\nu_N$ and $Q^\nu_E$ as well as the energy loss rate $Q$ are calculated in the same way as for the electron-positron pair 
annihilations by Eqs.~(\ref{eq:eq-dLN}), (\ref{number}), (\ref{LE}) and (\ref{energy}). The right panels of Fig.~{\ref{fig7}} show the number spectra for different $\rho Y_e$ (top panel) and temperatures 
(bottom panel). It is found from the figure that the number spectrum is much less sensitive to temperature than for the electron-positron pair annihilation but depends more on $\rho Y_e$. It is
also evident that the peak energy is considerably smaller in the plasmon decay compared with the pair annihilation although the amplitudes are not so different between them. This fact has an important 
implication for observability of the neutrinos emitted by these processes on the terrestrial neutrino detectors.

The fitting formula to the energy loss rate was provided by \cite{haft94} in the following form: 
\begin{equation}
Q_{\mathrm{plasma}} = \left( {C_V}^2+2{C^\prime_V}^2\right)Q_V.
\label{eq:Qplasma}
\end{equation}
In the above expression $Q_V$ is given as 
\begin{equation}
Q_V=3.00\times 10^{21}\lambda^9 \gamma^6 e^{-\gamma} \left(f_T+f_L\right)f_{xy},
\end{equation}
in which the following notations are employed:
\begin{eqnarray}
\gamma^2 & = & \frac{1.1095 \times 10^{11} \rho/\mu_e}{T^2 \left[1+\left(1.019\times 10^{-6} \rho/\mu_e\right)^{\frac{2}{3}}\right]^{\frac{1}{2}}}\\
f_T & = & 2.4+0.6\gamma^{\frac{1}{2}}+0.51\gamma+1.25\gamma^{\frac{3}{2}}\\
f_L & = & \frac{8.6\gamma^2+1.35\gamma^{\frac{7}{2}}}{225-17\gamma+\gamma^2}\\
x & = & \frac{1}{6}\left[17.5+\log_{10}\left(2\rho/\mu_e\right)-3\log_{10}T\right]\\
y & = & \frac{1}{6}\left[-24.5+\log_{10}\left(2\rho/\mu_e\right)+3\log_{10}T\right]\\
f_{xy} & = &
\left \{
\begin{array}{l}
1 \ \ \ \ \ \ \ \ \ \ \ \ \ \ \ \ \ \ \ \ \ \ \ \ \ \ \ \ \ \ \ \ \ \ \ \ \ \ \ \ \ \ \ \ \ \ \ \ \ \ \ \ \ \ \ \ \ \ \ \ \ \ \ \ \ \ \ \ \ \ \ \ \ \ \ \ \ \ \ \ \ \ \ (|x|>0.7\ \mathrm{or}\ y<0)\\ 
1.05 + \{ 0.39 -1.25x-0.35\sin{\left(4.5x\right)} \\
\ \ \ \ -0.3\exp{\left[-\left(4.5x+0.9\right)^2\right]} \} \times \exp{\left\{ -\left[\displaystyle{\frac{\mathrm{min}\left(0,y-1.6+1.25x\right)}{0.57-0.25x}}\right]^2 \right\}} \ \ \ \ \ (\mathrm{otherwise})
\end{array}
\right.
\end{eqnarray}
The error of this fitting formula is less than {5\ \%} in the regime, where the plasmon decay dominates in the cooling over other processes. We compared the energy loss rate obtained by our formula 
and that given by Eq.~(\ref{eq:Qplasma}) in the right panel of Fig.~\ref{fig8}, in which the temperature is fixed to $T=10^{10}\ {\rm K}$ just as in the pair annihilation case. Again the contribution from 
electron-type neutrinos is taken into account in this comparison. It is apparent that they agree
with each other excellently. Note that the plasmon decay is dominant at $\rho Y_e \gtrsim 10^{10}\ {\rm g/cm^3}$ (\citealt{itoh1996},  Fig.~\ref{fig2}).

\section{Results}

In the following we present the main results: the number and energy luminosities as well as the spectra for different neutrino flavors as functions of the time to collapse. Based on them, 
we then estimate the expected numbers of detection events for different terrestrial neutrino detectors. 

\subsection{luminosity and spectrum}

Since stellar cores are not homogeneous, we need to calculate the number and energy emissivities per volume and time, $Q_{N}^{\nu}$ and $Q_{E}^{\nu}$, as well as the number spectra,
$dQ_{N}^{\nu}/dE_{\nu}$, at each point of the core and integrate them with respect to radius to obtain the number and energy luminosities, $L_{N}^{\nu}$ and $L_{E}^{\nu}$, together with 
the observed number spectra, $dL_{N}^{\nu}/dE_{\nu}$: 
\begin{eqnarray}
L_{N}^{\nu} & = & \int_0^R  Q_{N}^{\nu}(r) \, 4\pi r^2 dr, \\
L_{E}^{\nu} & = & \int_0^R  Q_{E}^{\nu}(r) \, 4\pi r^2 dr, \\
\frac{dL_{N}^{\nu}}{dE_{\nu}} & = & \int_0^R \frac{dQ_{N}^{\nu}(r)}{dE_{\nu}} \, 4\pi r^2 dr.
\end{eqnarray}
Here $R$ denotes the radius of the core surface. We evaluate them at different times so that their time evolutions could be obtained.

Fig.~{\ref{fig9}} shows the number luminosities of different neutrino species for the three progenitor models as functions of the time to collapse. The contributions from the pair annihilation 
and the plasmon decay and their sum are plotted separately for the $8.4\ {\rm M_{\odot}}$ model whereas only those from the pair annihilation are shown for the other two progenitors. 
The left panel displays the results for the electron-type neutrinos and those for the other types are presented in the right panel. Since there is no heavy charged lepton in the progenitor
cores, there is no difference between the mu-type and tau-type neutrinos. To the production of electron-type neutrinos, on the other hand, not only the neutral current but the charged current 
also contribute. The number luminosities of electron-type neutrinos are hence larger by an order of magnitude than others in general. This is particularly true of the plasmon decay in the 
$8.4\ {\rm M_{\odot}}$ model. In fact, the number luminosity by the plasmon decay is larger than by the pair annihilation for the electron-type neutrinos whereas the former is much smaller 
than the latter for the mu- and tau-type neutrinos. The difference between the neutrino flavors, although not so drastic, are also recognized for the pair annihilation. 

Much more remarkable in the figure is qualitative differences between the ONe core and the Fe cores. Among other things, the Fe cores emit a substantial amount of neutrinos from much 
earlier on than the ONe core. This is simply due to their higher temperatures as mentioned earlier. The electron-positron pairs are more abundant, producing neutrinos copiously. In fact, the 
ONe core is so cold until the electron capture on Mg is opened that electrons are strongly degenerate and, as a consequence, positrons are scarce, making the neutrino luminosity from 
the pair annihilation negligibly small. Unfortunately, the plasmon contribution is also tiny as long as the temperature is very low, since only a small amount of plasmons are thermally populated then.
It is hence understandable that the neutrino luminosity becomes substantial in the ONe core only after the electron captures and the subsequent burnings of oxygen and silicon heat it up, which
occur less than a second before collapse. Even after the formation of the NSE region ($T \gtrsim 5 \times 10^9\ {\rm K}$) inside the core, the number luminosity of the ONe core is much smaller
than those of the Fe cores. As a matter of fact, $\sim 10^{53}$ neutrinos are emitted in the $15\ {\rm M_\odot}$ model while in the $8.4\ {\rm M_\odot}$ model the number of emitted neutrinos is 
only $\sim 10^{51}$. Note, however, that the luminosity increases very quickly in the last few hundred milliseconds in the ONe core and It is still increasing rapidly at the end of our calculation.
At this point the central density is $10^{11}\ {\rm g/cm^3}$ and no more data are available from the stellar evolution calculation. Unlike the Fe cores, the ONe core is rather slowly contracting
even at this point, with the NSE region being expanding gradually. We hence expect the luminosity continues to rise further until neutrinos are trapped inside the core, which will be somewhat 
delayed, since the neutrinos considered here have smaller energies ($\sim 5\ {\rm MeV}$) than those produced by electron captures during collapse ($\sim 10\ {\rm MeV}$).

We show the corresponding energy luminosity evolutions in Fig.~{\ref{fig10}}. Unlike the number luminosity, the energy luminosity of anti-neutrinos is different from that of their partner 
neutrinos for the electron-positron pair annihilation. They are hence displayed separately in the figure. In the case of the plasmon decay, however, the vector current contribution is dominant 
over other contributions and we ignore the latter completely in this paper. In that approximation, there is no distinction between the energy luminosity of neutrinos and that of the partner 
anti-neutrinos. It is apparent that the essential features are nearly the same as those in the number luminosities. The energy luminosities for the Fe cores are much larger than those for the 
ONe core through the entire evolutions up to collapse. Although the luminosities are different between the neutrinos and anti-neutrinos in the same flavor, differences are indiscernible in the
figure. One thing to note here is that the energy luminosity of $\nu_{e}$ from the plasmon decay is smaller than the that from the pair annihilation, the opposite trend to what we found in the 
number luminosity (see the left panel of Fig.~{\ref{fig9}}). This happens because the average energies of neutrinos are quite different between the two emitting processes, which will be
demonstrated shortly. Incidentally, both the number and energy luminosities are larger for the $15\ {\rm M_{\odot}}$ model than for the $12\ {\rm M_{\odot}}$ model. This is simply because the
temperatures are a bit higher for the former than for the latter according to the general rule that the more massive the star is, the hotter it is.  

Figure~{\ref{fig11}} shows the normalized number spectra of different neutrino species, which are evaluated for the pair annihilation and plasmon decay individually at different times for each 
of the three progenitor models. Note that upper panels correspond to earlier epochs in this figure. The results for the $8.4\ {\rm M_\odot}$ model are plotted in the left panels and are
limited to the last $\sim 200\ {\rm msec}$, since the luminosities are too small at earlier times as seen above. it is apparent from the figure that the peak neutrino energy for the plasmon 
decay is $\sim 0.1\ {\rm MeV}$ and is not much changed in this phase. These values are substantially lower than those for the pair annihilation. In fact, the neutrinos produced by the pair annihilation have $\sim 5-10\ {\rm MeV}$ on average. 
This is exactly the reason why the number luminosity of electron-type neutrinos from the plasmon decay is larger than that from the pair annihilation whereas the corresponding energy 
luminosities have the opposite order. 
 
It is also evident that the spectra for the pair annihilation are much broader than those for the plasmon decay and, as a consequence, a greater number of high energy neutrinos are emitted in the pair 
annihilation, the fact that makes this process more important from the observational point of view, since the cross section of the inverse $\beta$ decay process that is employed for detection 
by the terrestrial detectors is approximately proportional to the square of neutrino energy. We can also recognize that neutrinos have higher peak energies than anti-neutrinos of the same 
flavor. The difference is largest for the electron-type neutrino. The origin of these differences is the disparity between electrons and positrons in the cores, with the former being more abundant.
   
The middle and right panels in Fig.~{\ref{fig11}} show the normalized number spectra for the $12\ {\rm M_{\odot}}$ and $15\ {\rm M_{\odot}}$ models, respectively. In these plots the plasmon decay
is ignored and only those from the pair annihilation are displayed. Note that the time spans are quite different in these plots from that for the $8.4\ {\rm M_{\odot}}$ model. This is because 
the Fe cores emit neutrinos longer as mentioned earlier. What is most remarkable here is the fact that the peak energies for these Fe cores are much lower than those of the ONe core. 
As a matter of fact, the peak energies ($\sim 1 - 2\ {\rm MeV}$) for the former is even smaller than the threshold energy $E_{\rm th} \approx 1.8\ {\rm MeV}$ for the inverse $\beta$ decay.
The difference stems from the fact that electrons are more degenerate and have larger chemical potentials in the ONe core (see Fig.~\ref{fig3}). For the same reason, the difference
between the neutrinos and anti-neutrinos of the same flavor is much less pronounced in the Fe cores although the trend is the same. We will investigate what consequences the features
we have seen in this subsection may have on the detectability of the neutrinos emitted from the different progenitors in the next subsection.

\subsection{event numbers at detectors}

Based on the results obtained so far, we roughly evaluate the expected number of neutrinos that will be detected in the terrestrial detectors such as Super-Kamiokande and KamLAND. 
Both the water Cherenkov and liquid scintillation detectors will observe the inverse-$\beta$ decay reaction
\begin{equation}
\bar{\nu}_e + p \longrightarrow e^+ + n, 
\end{equation}
which is induced by the pre-supernova neutrinos when they hit the targets in the detectors. Following \citet{odrzy04}, we express the cross section $\sigma(E_\nu)$ of this interaction as 
\begin{equation}
\sigma(E_\nu)=\sigma_0\left({g_V}^2+3{g_A}^2\right)E_{e^+}p_{e^+}=0.0952\left(\frac{E_{e^+}p_{e^+}}{1\mathrm{MeV}^2}\right)\times 10^{-42} \ \ \ \ [\mathrm{cm}^2],
\end{equation}
in which the emitted positron energy and 3-momentum are denoted by $E_{e^+}=E_{\nu}-\left(M_n-M_p\right)$ and $p_{e^+}=\sqrt{{E_{e^+}}^2-{m_e}^2}$, respectively, with $M_n=939.6\ {\rm MeV}$, 
$M_p=938.27\ {\rm MeV}$ and $m_e=0.511\ {\rm MeV}$ being the neutron, proton and electron rest masses, respectively; the vector and axial-vector form factors are assumed to be $g_V=1$, 
$g_A=1.27$ and the constant $\sigma_0$ is given as 
\begin{equation}
\sigma_0=\frac{G^2\cos^2{\theta_c}}{\pi} \left(1+\Delta^R_{\mathrm{inner}}\right)
\end{equation}
with the radiative correction $\Delta^R_{\mathrm{inner}}\simeq0.024$ and the Cabbibo angle $\theta_c=0.974$. Then the event rate at a detector, $r$, is expressed as
\begin{equation}
r=\frac{N}{4\pi R^2}\int_{E_{\mathrm{th}}}^{\infty} dE_{\nu_1} \sigma\left(E_{\nu_1}\right) \frac{dL_{N}^{\nu_1}}{dE_{\nu_1}}, 
\label{eq:eq42}
\end{equation}
in which $N$ and $R$ denote the target number in the detector and the distance between the star and the detector, respectively. For simplicity, we assume that the detection efficiency is $100\ \%$
above the threshold $E_{\rm th}$. The features we assume in this paper are summarized in Table~\ref{detector} for Super-Kamiokande, KamLAND, Hyper-Kamiokande and JUNO. For the latter two,
the actual numbers will be different\footnote{For JUNO, as a matter of fact, we just scale KamLAND by a factor of 20.} from those listed in the table. Since it is not our intention here to give very 
accurate estimates on the event numbers, we believe that they are good enough. The cumulative event number $N_{\rm cum}$ is obtained by the integration over the time:
\begin{equation}
N_{\rm cum}(t) = \int_{t_{\rm ini}}^t r \, dt.
\end{equation}

In evaluating Eq.~(\ref{eq:eq42}), the number spectrum should be modified with an appropriate account of neutrino oscillations. We show in Fig.~\ref{fig17} the mixing lengths and the density
scale heights along with the density profiles for the three progenitor models. We assume here that the neutrino energy is $5\ {\rm MeV}$ and that the mass hierarchy is inverted. Since the 
envelope of the $8.4\ {\rm M_{\odot}}$ is much more tenuous than those for the other two models and the density gradient is much steeper accordingly, the ratio of the scale height to the 
mixing length becomes smallest for the ONe core model at the resonance, which is marked with a star in the figure, just as expected. It is evident, however, that the scale height is still larger 
than the mixing length by a order of magnitude even in that case. We hence assume adiabatic oscillations in the following. This is certainly justified for the normal hierarchy, since there is no
resonance in the anti-neutrino sector. It is noted, however, that the mass and structure of the outer envelope of the 8.4$\ {\rm M}_{\odot}$ model are highly uncertain just prior to collapse. Since the envelope was ignored in the evolution calculation after the completion of C burning, we attach it again to the core model, assuming that it is hydrostatic and its temperature and chemical composition profiles are unchanged from those at the end of C burning.

The number spectrum of the electron-type anti-neutrino is then given as follows:
\begin{equation}
\left( \frac{dL_{N}^{\bar{\nu}_e}}{dE_{\bar{\nu}_e}}\right)_{\rm osc} = p \left( \frac{dL_{N}^{\bar{\nu}_e}}{dE_{\bar{\nu}_e}}\right)_0 
+ (1 - p)\left( \frac{dL_{N}^{\bar{\nu}_x}}{dE_{\bar{\nu}_x}}\right)_0. 
\end{equation}
In this expression, the subscript $0$ means the original spectra before the neutrino oscillations are taken into account; $\bar{\nu}_{x}$ stands for $\bar{\nu}_{\mu}$ or $\nu_{\tau}$, both of which
we assume to have the same spectra; the so-called survival probability $p$ is given in the adiabatic limit as
\begin{equation}
p = \left\{\begin{array}{l}
|U_{e1}|^2 = \cos^2\theta_{12} \cos^2 \theta_{13}\quad \mbox{for normal hierarchy,}\\
|U_{e3}|^2 = \sin^2 \theta_{13}\quad \mbox{for inverted hierarchy.}
\end{array} \right.
\end{equation}
with $\cos^2{\theta}_{12}=0.692$,  $\cos^2{\theta}_{13}=0.9766$ for normal hierarchy and $\sin^2{\theta}_{13}=0.024$ for inverted hierarchy \citep{pdg}.
In the above expression, the $U_{e1}$ and $U_{e3}$ are the elements of the unitary matrix $U$ transforming the mass eigen states to the flavor eigenstates as
\begin{equation}
\left(\begin{array}{c} \nu_e \\ \nu_{\mu} \\ \nu_{\tau} \end{array} \right) = U \left(\begin{array}{c}\nu_1 \\ \nu_2 \\ \nu_3 \end{array} \right).
\end{equation} 
It is normally parametrized by three mixing angles, $\theta_{12}$, $\theta_{13}$, $\theta_{23}$ and a CP-violating phase, $\delta$, as 
\begin{equation}
U = \left( \begin{array}{ccc} U_{e1} & U_{e2} & U_{e3}\\ U_{\mu 1} & U_{\mu 2} & U_{\mu 3}\\ U_{\tau 1} & U_{\tau 2} & U_{\tau 3} \end{array} \right)
= \left( \begin{array}{ccc} c_{12}c_{13} & s_{12}c_{13} & s_{13}e^{-i\delta}\\ -s_{12}c_{23} - c_{12}s_{13}s_{23} e^{i\delta} & c_{12}c_{23} - s_{12}s_{13}s_{23} e^{i\delta} & c_{13}s_{23}\\ 
s_{12}s_{23} - c_{12}s_{13}c_{23}e^{i\delta} & - c_{12}s_{23} - s_{12}s_{13}c_{23}e^{i\delta} & c_{13}c_{23} \end{array} \right)
\end{equation}
with the common shorthand notations of $c_{ij} = \cos \theta _{ij}$ and $s_{ij} = \sin \theta _{ij}$ for $i, j = 1, 2, 3$.

Fig.~{\ref{fig14}} show the cumulatice event numbers at Super-Kamiokande and KamLAND as functions of time. The upper panel is for the Fe core models whereas the lower one 
corresponds to the ONe core model. The distance to the star is assumed to be $200 \ {\rm pc}$, the estimated distance to Betelgeuse. In the case of the Fe core models, more than 
1 neutrinos will be detected a day before collapse both at Super-Kamiokande and KamLAND, which seems to be consistent with the preceding paper \citep{odrzy04}. The total 
event numbers are also comparable between the two detectors although the detector volumes are vastly different. This is mainly due to the sensitivity of KamLAND to low energy
neutrinos (see Table~1 for the energy threshold of each detector). 

In the case of the ONe core model, on the other hand, the expected event numbers are much smaller than those for the Fe core models. This is so in spite of the fact that
the peak energy of $\bar{\nu}_e$ is about twice larger for the ONe core than for the Fe cores after the formation of NSE core in the former. It turns out that the duration of 
such neutrinos is simply too short in the ONe core. Since the volume of the NSE core is small accordingly, the luminosity itself is smaller for the ONe core. 

Both the water Cherenkov detector and liquid scintillation detector may see a major scale-up in the near future. For example, Hyper-Kamiokande is a planned next-generation 
water Cherenkov detector with volume expected to be $\sim 15$ times larger than that of Super-Kamiokande. JUNO, on the other hand, is a liquid scintillation detector with 
the size $\sim 20$ times as great as that of KamLAND ans is currently under construction. Simply assuming the parameters summarized in Table~1, which will certainly 
different from the actual numbers one way or another, and ignoring the detection efficiencies again, we calculated the event numbers for these future detectors. The cumulative event 
numbers are shown in Fig.{\ref{fig16}} as functions of time. The distance to the source is again assumed to be $200\ {\rm pc}$. 

It is found that for the $15\ {\rm M_{\odot}}$ model with an Fe core, the total event number reaches $\sim 266 - 864$ for JUNO, which means that we may observe $\sim 10 - 35$ 
events even if the source is located at $1\ {\rm kpc}$. Hyper-Kamiokande, on the other hand, will observe $\sim 28 - 77$ events, which is not much different from those for Super-Kamiokande. 
This is because of the somewhat high energy threshold of $8.3\ {\rm MeV}$ we employ here. If this were as small as the one for Super-Kamiokande, the event number would be larger 
by more than an order of magnitude. The disadvantage of the high energy threshold is also vindicated in the expected event numbers for the $12\ {\rm M_{\odot}}$ model. Since the 
peak energy of neutrino is somewhat lower in this model, the event numbers at Hyper-Kamiokande is even smaller than those at Super-Kamiokande. In the case of the 
$8.4\ {\rm M_\odot}$ model with an ONe core, the event numbers are expected to be much smaller than unity even for these large scale detectors and the distance as small as 
$200\ {\rm pc}$. We are hence forced to conclude that it is highly difficult to observe the pre-supernova neutrinos
from ONe cores. Put another way, no detection of such neutrinos from a nearby CCSN in the future may indicate that it is an ECSN.

\section{summary and discussions}
In this paper, we have investigated the pre-supernova neutrino emissions from the ONe-core forming progenitor and compared them with those from the more massive Fe-core producing progenitors,
aiming to distinguish ECSNe form more common FeCCSNe by future observations of a near by supernova. Employing the realistic progenitor models, we have calculated the temporal 
evolutions of the luminosities and spectra of the neutrinos emitted by these stars via the electron-positron pair annihilations and plasmon decays. Based on these results, we have then 
roughly evaluated the numbers of detection events for Super-Kamiokande and KamLAND as the representative water Cherenkov and liquid scintillation detectors, respectively, in current 
operation. We have also considered Hyper-Kamiokande and JUNO, which may be operational in the near future. 

We have found that the electron-positron pair annihilation dominates over other neutrino-emission processes in the Fe-core producing models whereas in the ONe-core generating model 
the plasmon decay is dominant until the NSE core is formed at the center, after which the pair annihilation prevails quickly and the neutrino luminosity rises rapidly. Since the contraction
proceeds slowly in the latter case, the central density is larger just prior to collapse and, as a result, emitted neutrinos have larger peak energies than in the Fe cores, which is advantageous 
from the observational point of view. It is also found that the neutrinos emitted via the plasmon decay have much smaller energies ($\sim 0.1\ {\rm MeV}$) and will be undetected on the 
terrestrial detectors that employ the inverse decay reaction with the energy threshold of $\approx 1.8\ {\rm MeV}$.
  
Based on these results we have roughly estimated the number of detection events for Super-Kamiokande and kamLAND. We have seen that both will detect tens of neutrinos from the 
FeCCSNe if they are located at $200\ {\rm pc}$ from the earth whereas neither will be able to detect the ECSNe even at this small distance. If we envisage next-generation detectors such 
as JUNO, these numbers may be increased by an order of magnitude. On the other hand, Hyper-Kamiokande may not be advantageous for the detection of such low energy neutrinos
unless the energy threshold is as low as that of Super-Kamiokande. 

It should be pointed out that the neutrino emissions will continue after the gravitational collapse of the core commences. As mentioned earlier, the core contraction is slow in the 
$8.4\ {\rm M_{\odot}}$ model even when the central density reaches $10^{11}\ {\rm g/cm^3}$ and it may take another few hundreds msec until core bounce. Since the neutrino 
luminosity is till rising rapidly even at the end of calculation. It is hence expected that the actual event numbers will be increased substantially. Some delay of the neutrino trapping, which
will actually terminate the neutrino emissions, owing to the lower neutrino energy will be also advantageous in this respect. It is hence important to calculate the neutrino light curve 
up to core bounce, after which ordinary supernova neutrinos will prevail. 

\begin{figure}
\plotone{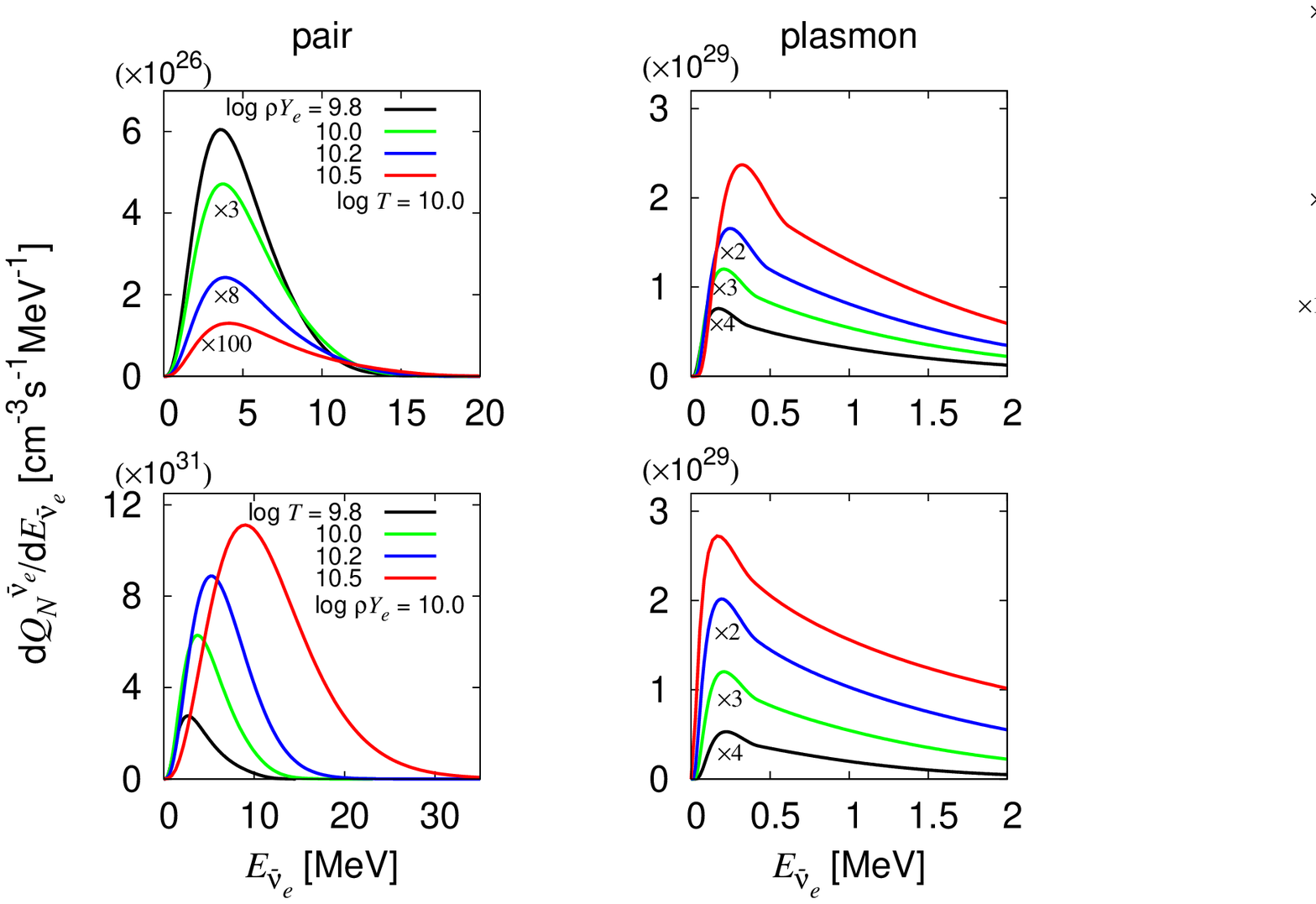}
\caption{The number spectra of the pair annihilation (left panels) and the plasmon decay (right panels) for different combinations of $\rho Y_e$ and $T$. The top panels show the dependence on $\rho Y_e$, in which T is fixed to $10^{10}\ $ K, whereas the lower panels display the dependence on T, where $\rho Y_e$ is set to $10^{10}\ {\rm g/cm^3}$.  \label{fig7}}
\end{figure}

\begin{figure}
\plottwo{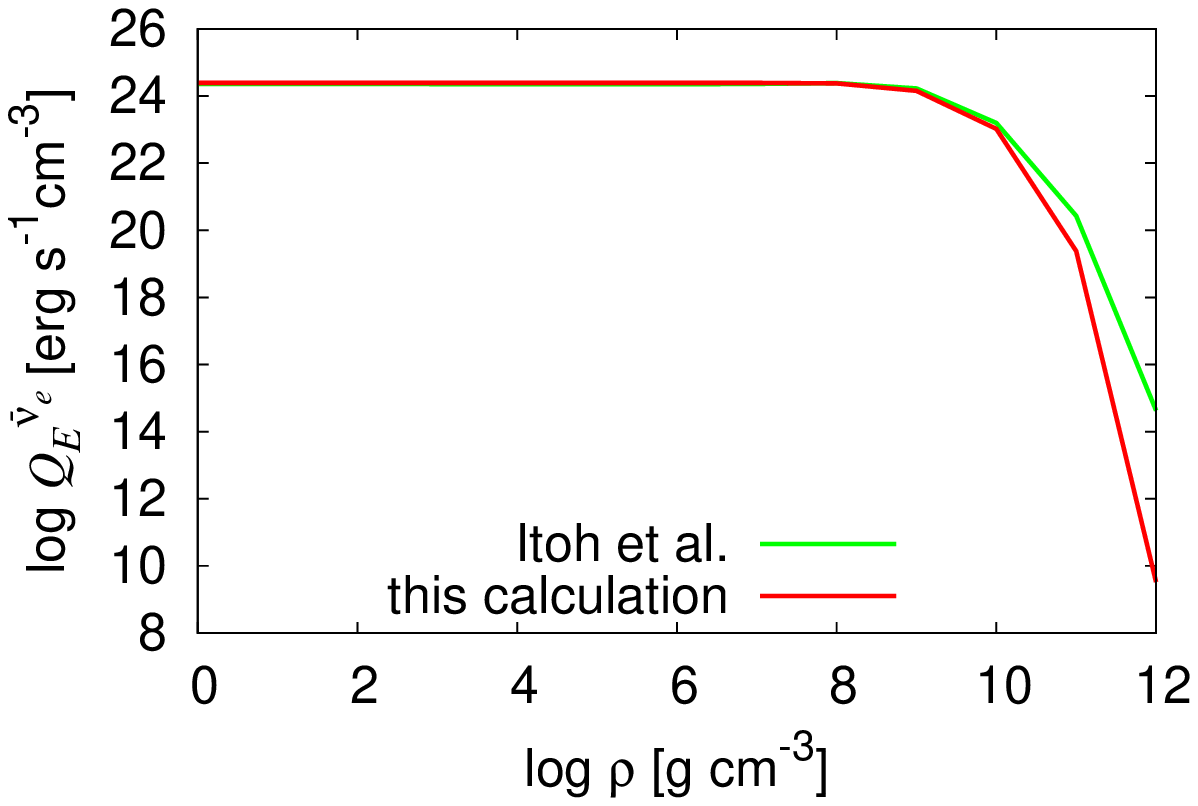}{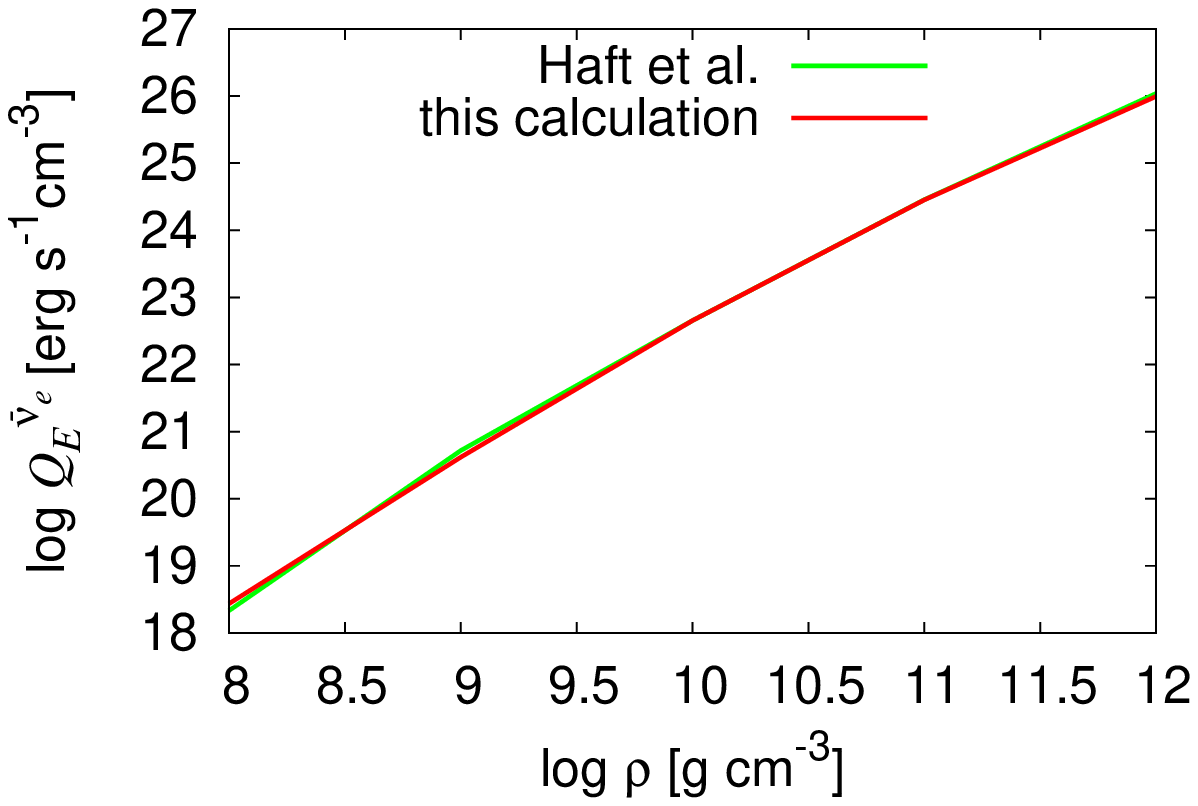}
\caption{Comparisons with the fitting formulae for the pair annihilation (left panel) and plasmon decay (right panel). The temperature and electron fraction are fixed to $10^{10}\ {\rm K}$ and $Y_e=0.5$, respectively, in both panels. \label{fig8}}
\end{figure}

\begin{figure}
\plotone{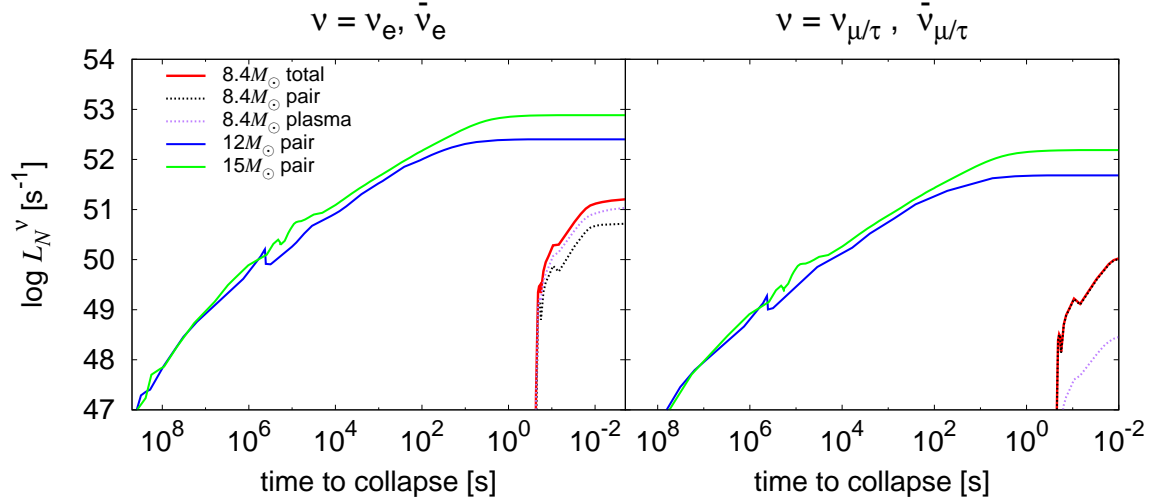}
\caption{The time evolutions of the number luminosities $L^\nu_{N}$ of electron-type neutrinos (left panel) and of other types (right panel) for three progenitor models. \label{fig9}}
\end{figure}

\begin{figure}
\plotone{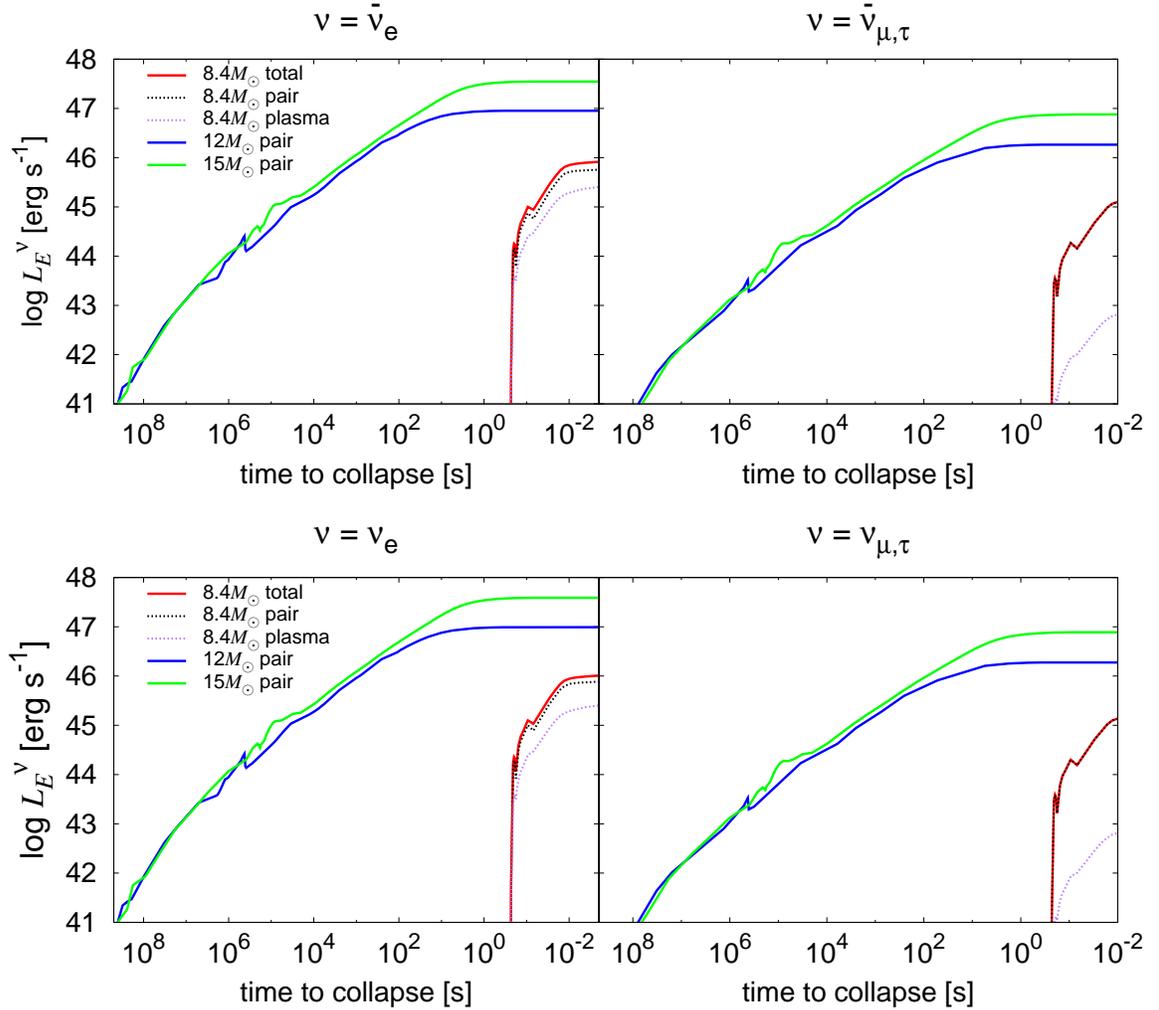}
\caption{The time evolutions of the energy luminosities $L^\nu_{E}$ of electron-type neutrinos (left panel) and of other types (right panel) for three progenitor models. Top and bottom panels show the results for neutrinos  and anti-neutrinos, respectively. \label{fig10}}
\end{figure}

\begin{figure}
\plotone{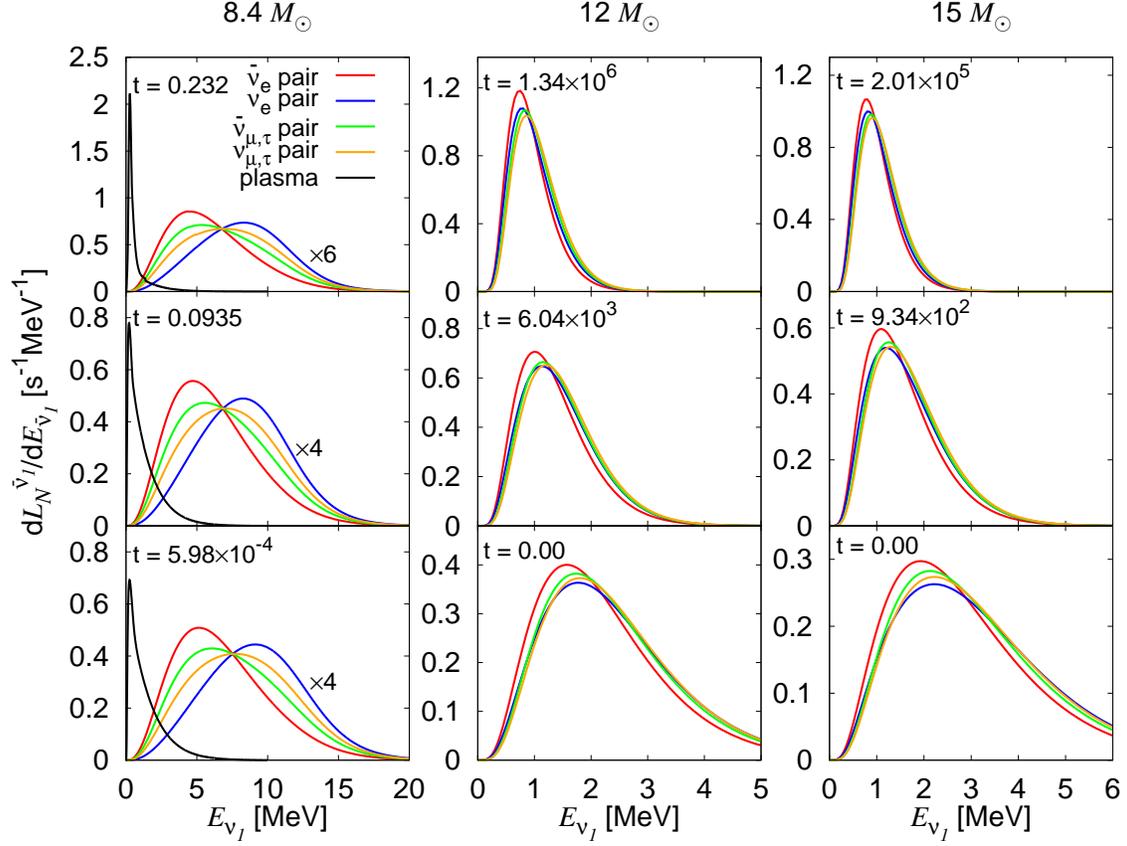}
\caption{Normalized number spectra at different times for $8.4\ {\rm M_{\odot}}$ (left panels), $12\ {\rm M_{\odot}}$ (middle panels) and $15\ {\rm M_{\odot}}$ (right panels). Red, blue, green and orange curves correspond, respectively, to $\bar{\nu}_{e}$ and $\nu_{e}$ from the pair annihilation and $\bar{\nu}_{\mu}/\bar{\nu}_{\tau}$ and $\nu_{\mu}/\nu_{\tau}$ from the pair annihilation. All neutrinos have the identical spectrum after normalization for the plasmon decay as shown with black. For better visibility, all the lines but the black one in the left panels are multiplied by the factors indicated. Note that larger $t$'s correspond to earlier times. \label{fig11}}
\end{figure}

\begin{figure}
\plotone{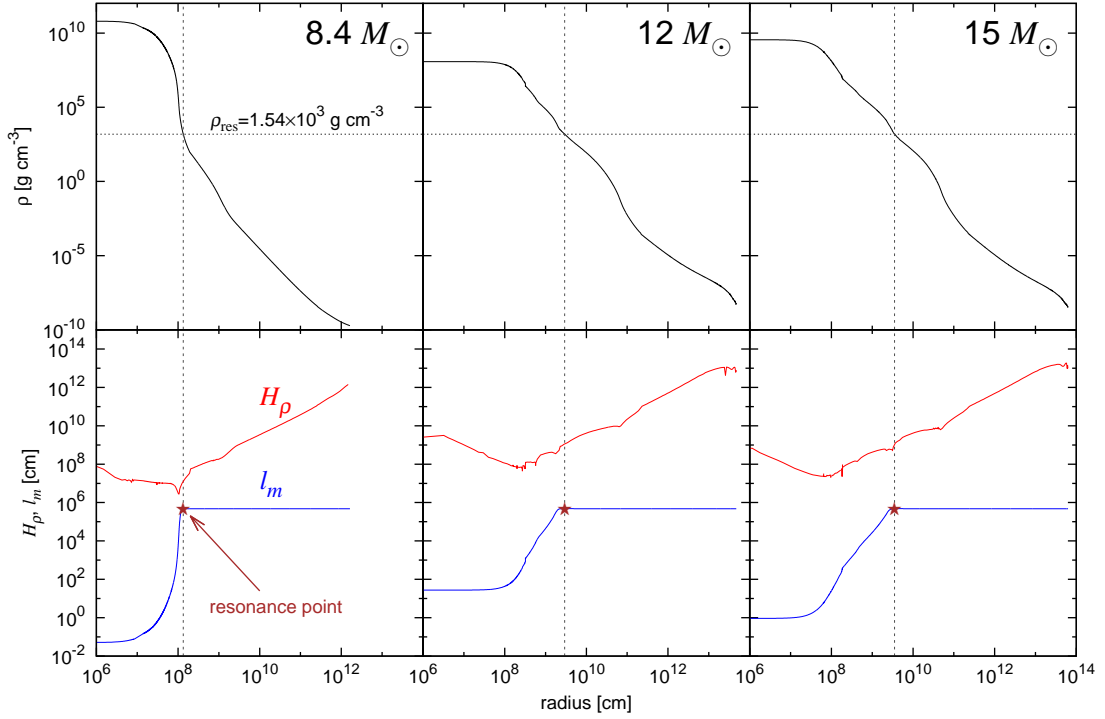}
\caption{Comparisons between density scale heights ($H_\rho$) and mixing lengths ($l_m$) for three progenitor models (lower panels). The inverted mass hierarchy is assumed and the resonance points are marked by star symbols and vertical lines. The radii of the resonance points for the 8.4, 12, 15$\ {\rm M_\odot}$ models are 1.3 $\times10^8$, 2.95$\times10^9$ and 3.53$\times10^9$ cm, respectively. In the upper panels we show the density profiles for convenience. \label{fig17}}
\end{figure}

\begin{figure}
\plottwo{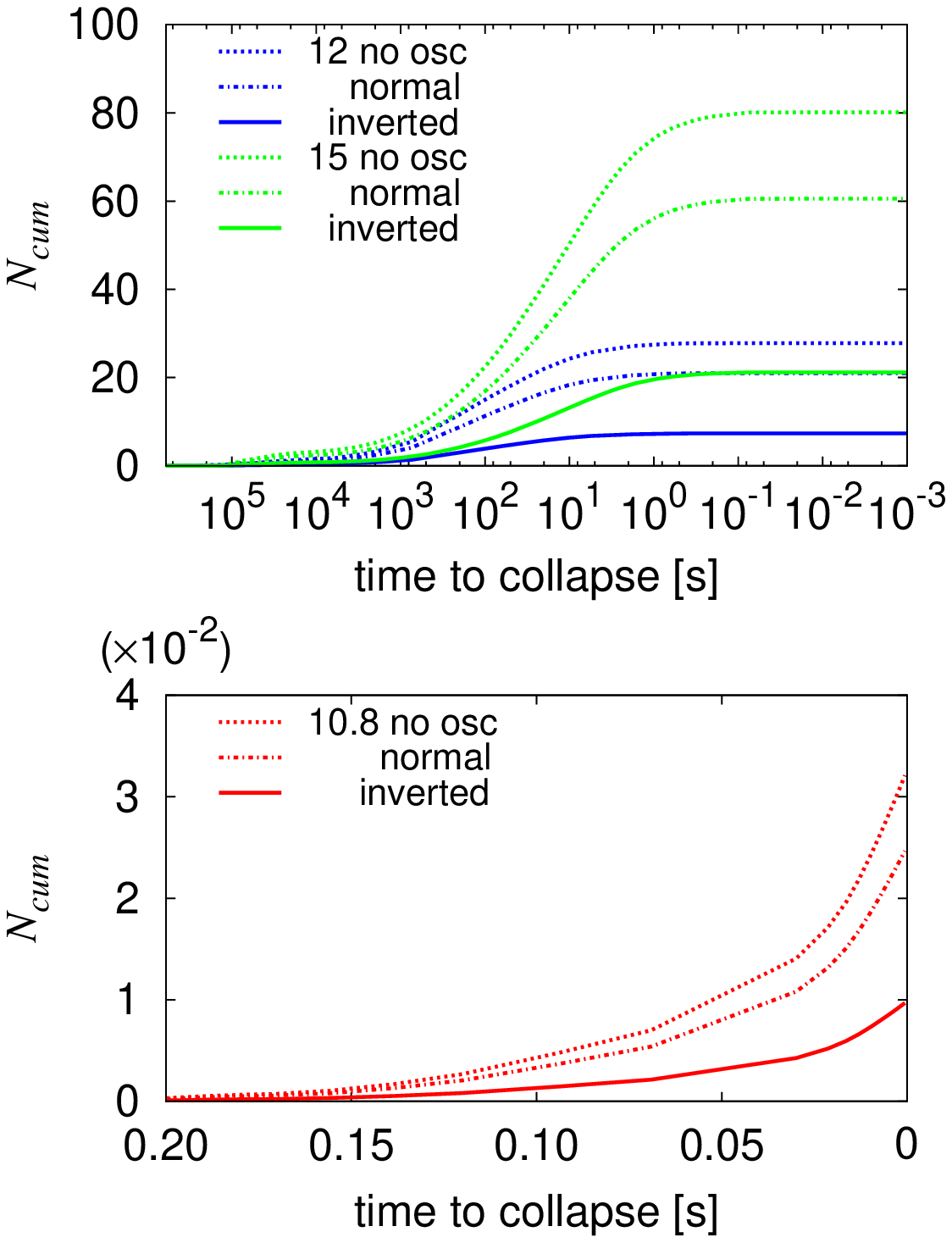}{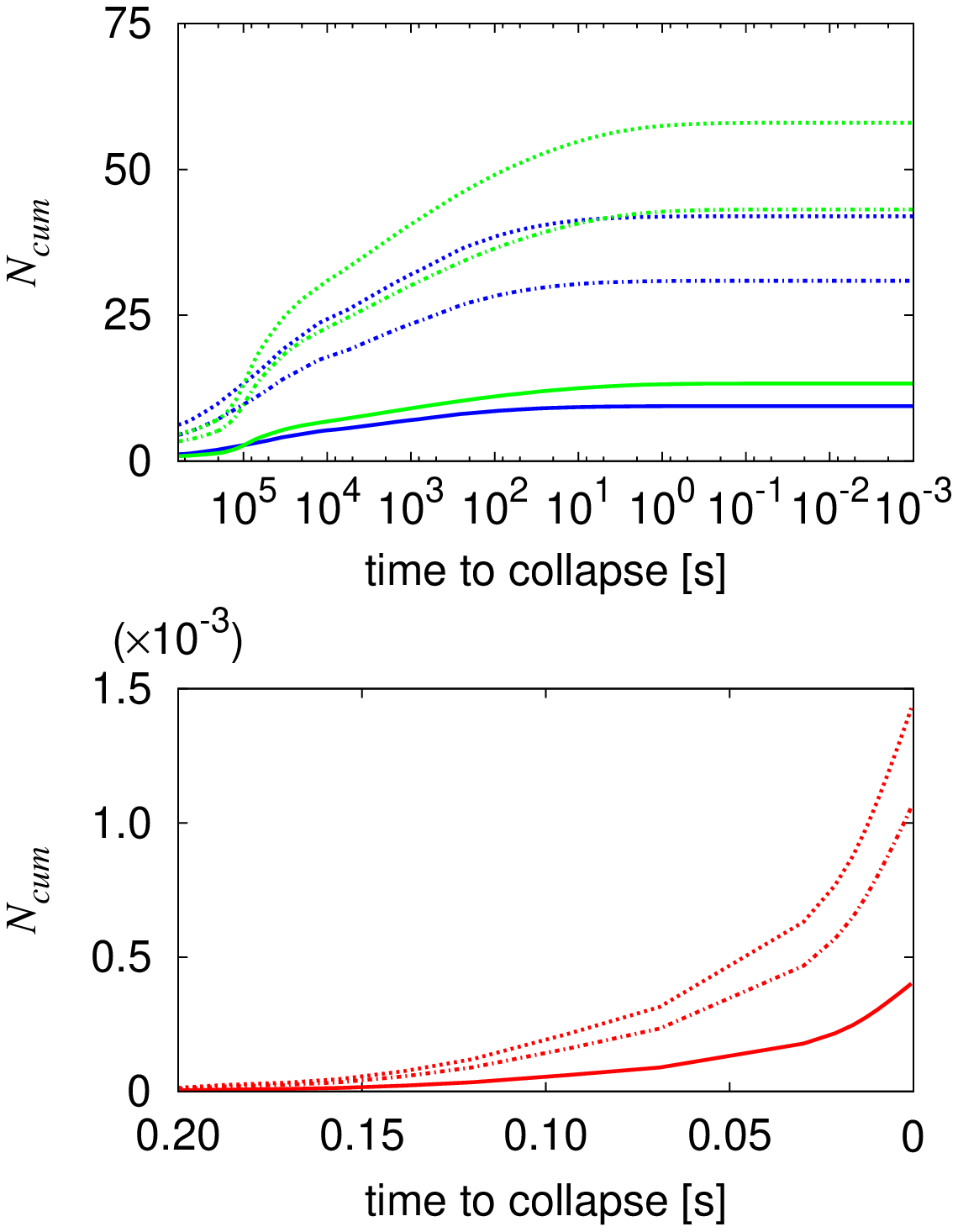}
\caption{The cumulative numbers of detection events at Super-Kamiokande (left panels) and KamLAND (right panels) as functions of time to collapse. The top panels show the results for the Fe-core models whereas the bottom panels display the evolutions for ONe-core model. Note the difference in the scales of the horizontal axis between the upper and lower plots. Red, blue and green lines correspond to the $8.4\ {\rm M_{\odot}}$, $12\ {\rm M_{\odot}}$ and $15\ {\rm M_{\odot}}$ models, respectively and solid, dotted and dashed curves show the results for no neutrino oscillation and those for adiabatic oscillations with the normal and inverted mass hierarchies, respectively. \label{fig14}}
\end{figure}

\begin{figure}
\plottwo{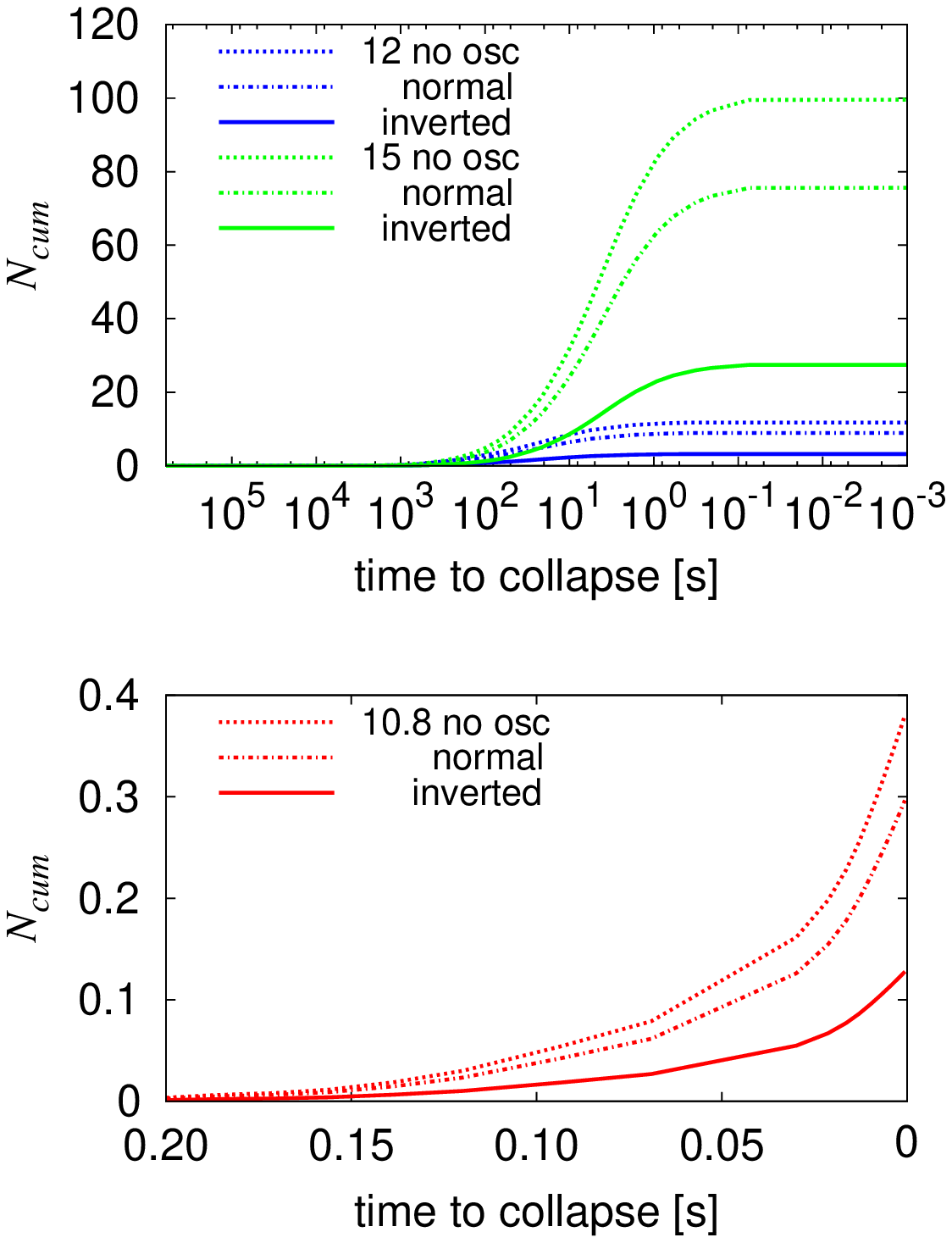}{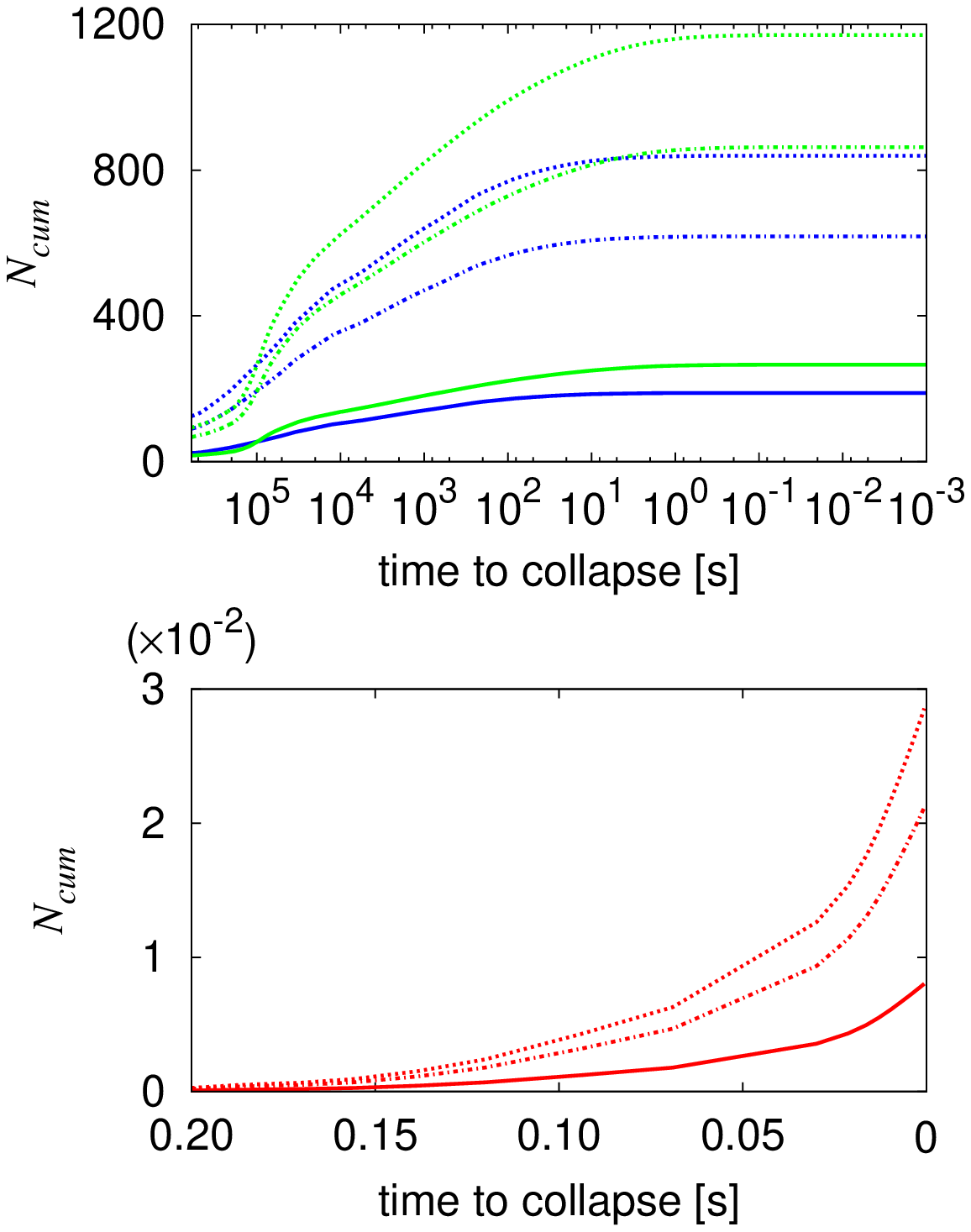}
\caption{The same as Fig.14 but for Hyper-Kamiokande (left panels) and JUNO (right panels). \label{fig16}}
\end{figure}

\begin{figure}
\plotone{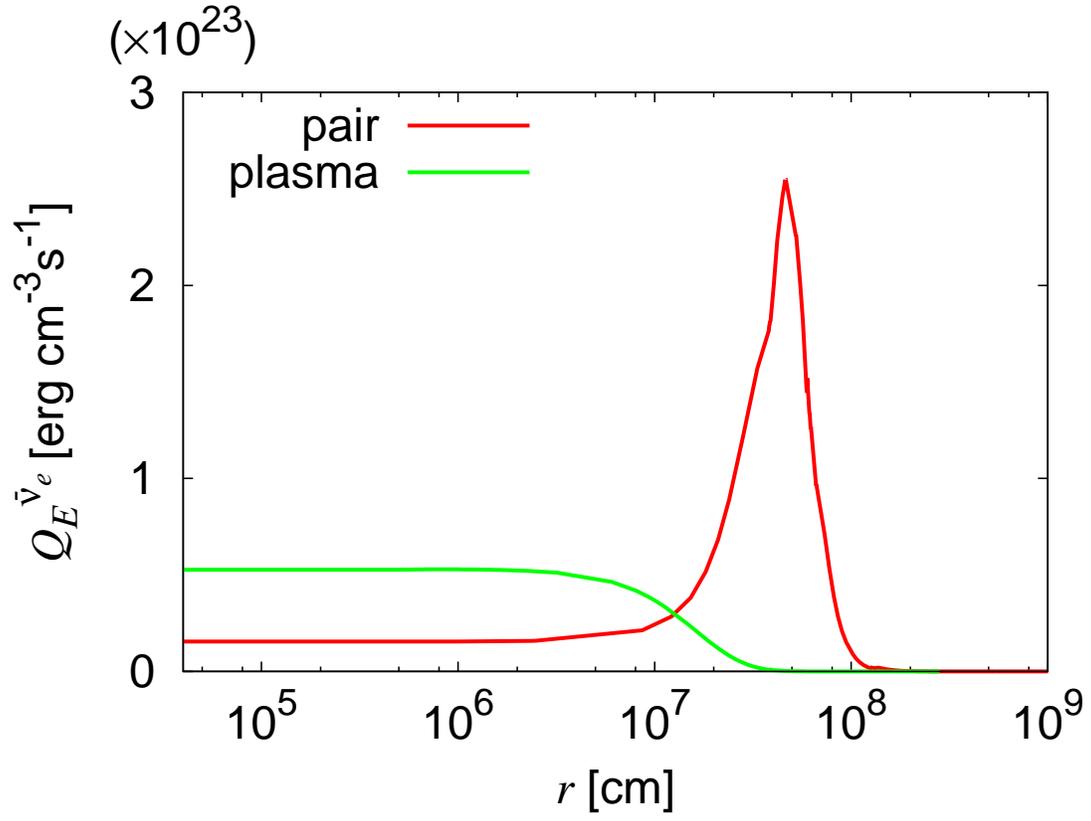}
\caption{Radial profiles of the neutrino emissivities for the pair annihilation and the plasmon decay in the $15\ {\rm M_{\odot}}$ just prior to collapse. \label{fig20}}
\end{figure}

Although we have not considered the plasmon decay for the Fe cores in this paper, it may be comparable to or even dominant over the pair annihilation in the late phase according to 
the right panel of Fig.~{\ref{fig2}}. We hence compare the total emissivities of these two processes just prior to core-collapse for the $15\ {\rm M_\odot}$ model in Fig.~{\ref{fig20}}. 
In the central region $r \lesssim 10^7\ {\rm cm}$, the plasmon process emits more energy indeed. In the outer part of  the core, however, the pair annihilation becomes dominant and 
emit far greater amounts of energy in neutrinos. This is due to the strong dependence of the pair annihilation on the temperature (see the lower left panel of Fig.~\ref{fig7}). Note also 
that the plasmon decay is suppressed at lower densities (see the upper right panel of Fig.~\ref{fig7}). Integrating these emissivities over the radius, we find that the total emitted energy 
from the plasmon process is $\sim 10^{45}\ {\rm erg/s}$, which is much smaller than that from the pair annihilation ($\sim 10^{47}\ {\rm erg/s}$) and justifies our neglect of the plasmon decay in 
the Fe-core progenitor models. 

So far we have seen that the pre-supernova neutrino emission is qualitatively different between the ONe and Fe cores, which fact is reflected in the expected event numbers on the terrestrial
detectors. Actual detectability is crucially dependent on the background noises at detectors, however. Although it is much beyond the scope of this paper to take them into account in detail
for each detector and discuss the detection possibility quantitatively, we can touch the issue briefly: if we adopt several hundreds$\ {\rm events/day}$ as a typical background for Super-Kamiokande, 
neutrinos from FeCCSNe may be detected at $3\sigma$ if the source is located at $200\ {\rm pc}$. The background will be reduced substantially if the neutron tagging is enhanced with Gd \citep{beacom04}. 
The background at KamLAND is low $\sim 1\ {\rm event/day}$. The neutrinos from Fe cores may be detected to ${\rm kpc}$. For the next-generation detectors, the distance may be
further extended. In the case of Hyper-Kamiokande, however, the reduction of the energy threshold will be more crucial as mentioned above. In this paper, we have considered only two relatively 
light Fe-core progenitors. It is certainly important, though, to study more systematically pre-supernova neutrinos from different progenitors. It should be also stressed that the expected event numbers for these models may change by a factor of a few if one considers various uncertainties in the stellar evolution calculation. These issues should be investigated more in detail and are actually being currently undertaken (\citealt{yoshida} in preparation).

\appendix

In the following we give detailed derivations of the expressions for the reaction rates of the electro-positron pair annihilation and
the plasmon decay in turn.

\section{reaction rate of electron-positron pair annihilation}
The rate of the reaction, $R$, to produce a pair of neutrino and anti-neutrino via the annihilation of a pair of electron and positron, 
which corresponds to the left Feynman diagram in Fig.~{\ref{fig0}} to the lowest order, is given by the low-energy limit of the Weinberg-Salam theory, 
which is actually identical to the Fermi's theory, as follows:
\begin{eqnarray}
&&R=\left(\frac{G}{\sqrt{2}}\right)^2\int\!\!\!\int\frac{d^3k}{\left(2\pi\right)^32E_e}\frac{d^3k^{\prime}}{\left(2\pi\right)^32{E_e}^\prime}
\left(2\pi\right)^4\delta^4\left(q+q^{\prime}-k-k^{\prime}\right)f_{e^-}\left(E_e\right)f_{e^+}\left(E_e^\prime\right) \nonumber \\ 
&&\times64\left[\left(C_V-C_A\right)^2\left(q\cdot k\right)\left(q^{\prime}\cdot k^{\prime}\right)+\left(C_V+C_A\right)^2
\left(q\cdot k^{\prime}\right)\left(q^{\prime}\cdot k\right)+m_e\left(C_V^2-C_A^2\right)\left(q\cdot q^{\prime}\right) \right],  \nonumber \\
&&
\end{eqnarray}
in which $k=(E_e,\mbox{\boldmath$k$}) $, $k^\prime=({E_e}^\prime,\mbox{\boldmath$k$}^\prime)$, $q=(E_\nu,\mbox{\boldmath$q$})$ and $q^\prime=(E_{\bar{\nu}},\mbox{\boldmath$q$}^\prime)$ 
denote the four momenta of electron, positron, neutrino and anti-neutrino, respectively. Following \cite{schinder1982,mezza1993}, we re-cast the above equation into the following form: 
\begin{equation}
R=\displaystyle{\frac{8G^2}{\left(2\pi\right)^2}\left[\beta_1I_1+\beta_2I_2+\beta_3I_3\right]},
\end{equation}
in which the factors are given as $\beta_1=\left(C_V-C_A\right)^2,\beta_2=\left(C_V+C_A\right)^2,\beta_3=C_V^2-C_A^2$ and the integrals are grouped into
\begin{eqnarray}
I_1&=&\int\!\!\!\int\frac{d^3k}{E_e}\frac{d^3k^{\prime}}{{E_e}^\prime}\delta^4\left(q+q^{\prime}-k-k^{\prime}\right)f_{e^-}\left(E_e\right)f_{e^+}\left(E_e^\prime\right) \left(q\cdot k\right)^2, \\
I_2&=&\displaystyle{\int\!\!\!\int\frac{d^3k}{E_e}\frac{d^3k^{\prime}}{{E_e}^\prime}\delta^4\left(q+q^{\prime}-k-k^{\prime}\right)f_{e^-}
\left(E_e\right)f_{e^+}\left(E_e^\prime\right) \left(q\cdot k^{\prime}\right)^2}, \\
I_3&=&\displaystyle{\int\!\!\!\int\frac{d^3k}{E_e}\frac{d^3k^{\prime}}{{E_e}^\prime}\delta^4\left(q+q^{\prime}-k-k^{\prime}\right)f_{e^-}
\left(E_e\right)f_{e^+}\left(E_e^\prime\right)  m_e^2\left(q\cdot q^{\prime}\right)^2 }.
\end{eqnarray}

These integrals are evaluated as follows. We begin with $I_1$. Using three of the four $\delta$-functions we can easily accomplish the integrals over the positron momentum $k^\prime$ to
get 
\begin{equation}
\int\frac{d^3k^{\prime}}{{E_e}^\prime}f_{e^+}\left(E_e^\prime\right)\delta^4\left(q+q^{\prime}-k-k^{\prime}\right)=
2f_{e^+}\left(E_{\nu}+E_{\bar{\nu}}-E_e\right)\Theta\left(E_{\nu}+E_{\bar{\nu}}-E_e\right)\delta\left({q+q^{\prime}-k}^2-m_e^2\right),
\end{equation}
in which the Heaviside function is denoted by $\Theta$. The remaining integrals over the electron momentum $k$ are performed on the spherical coordinates in the momentum space with 
the volume element written as $d^3k=|\mbox{\boldmath$k$}| E_e dE_e d(\cos{\theta_e}) d\phi_e$. The $\phi_e$ integral is trivial to give a factor of $2\pi$. The integral over $\theta_e$ can
be accomplished with the use of the last $\delta$-function. The resultant expression is given as 
\begin{eqnarray}
I_1=\int_{E_{\mathrm{min}}}^{E_\mathrm{max}} \!\! 2\pi \, dE_e \, f_{e^-}\left(E_e\right)f_{e^+}\left(E_{\nu}+E_{\bar{\nu}}-E_e\right)
\frac{{E_{\nu}}^2{E_{\bar{\nu}}}^2}{\Delta_e^5}\left(1-\cos{\theta}\right)^2 \left[AE_e^2+BE_e+C
\right], 
\end{eqnarray}
in which $\theta$ is the angle between {\boldmath $q$} and {\boldmath $q$}$^\prime$ and ${\Delta_e}^2$ is given by
\begin{equation}
{\Delta_e}^2\equiv{E_{\bar{\nu}}}^2+{E_{\nu}}^2+2E_{\nu}E_{\bar{\nu}}\cos{\theta}, 
\end{equation}
and $A$, $B$ and $C$ are defined as
\begin{eqnarray}
&\left \{
   \begin{array}{l}
       A\equiv {E_{\bar{\nu}}}^2+{E_{\nu}}^2-E_{\nu}E_{\bar{\nu}}\left(3+\cos{\theta}\right), \\
       B\equiv-2{E_{\nu}}^2+{E_{\bar{\nu}}}^2\left(1+\cos{\theta}\right)+E_{\nu}E_{\bar{\nu}}\left(3-\cos{\theta}\right),\\
       C\equiv\displaystyle{\left(E_{\nu}+E_{\bar{\nu}}\cos{\theta}\right)^2-\frac{1}{2}{E_{\bar{\nu}}}^2\left(1-\cos{\theta}^2\right)-\frac{1}{2}\left(\frac{m_e\Delta_e}{E_{\nu}}\right)^2\frac{1+\cos{\theta}}{1-\cos{\theta}} }.
   \end{array}
\right . 
\end{eqnarray}
The lower and upper limits of the remaining integral with respect to $E_e$ are given as
\begin{eqnarray}
\left \{
  \begin{array}{l}
      \displaystyle{E_{\mathrm{min}}=\mathrm{max}\left [m_e,\frac{E_{\nu}+E_{\bar{\nu}}}{2}-\frac{\Delta_e}{2}\sqrt{1-\frac{2{m_e}^2}{E_{\nu}E_{\bar{\nu}}\left(1-\cos{\theta}\right)}}\ \right]}, \\
      \displaystyle{E_{\mathrm{max}}=\mathrm{min}\left [E_{\nu}+E_{\bar{\nu}}-m_e,\frac{E_{\nu}+E_{\bar{\nu}}}{2}+\frac{\Delta_e}{2}\sqrt{1-\frac{2{m_e}^2}{E_{\nu}E_{\bar{\nu}}\left(1-\cos{\theta}\right)}}\ \right]}.
   \end{array}
\right .
\end{eqnarray}

Finally, we express the integral over the electron energy $E_e$ with the Fermi integral defined as
\begin{equation}
F_n\left(\eta \right)=\int_{0}^{\infty}\frac{x^n}{e^{x-\eta}+1}dx.
\end{equation}
Using the following relation for the product of the Fermi-Dirac distributions, 
\begin{eqnarray}
f_{e^-}\left(E_e\right)f_{e^+}\left(E_{\nu}+E_{\bar{\nu}}-E_e\right) & = & \displaystyle{\frac{1}{\exp[(E_{\nu}+E_{\bar{\nu}})/T]-1}}\nonumber \\
&\times &\left\{\displaystyle{\frac{1}{\exp[(E_e-\left(E_{\nu}+E_{\bar{\nu}}\right)-\mu_e)/T]+1}
-\frac{1}{\exp[(E_e-\mu_e)/T]+1} } \right\}, \nonumber \\
&&
\end{eqnarray}
we can obtain the final expression for $I_1$ as
\begin{eqnarray}
&&I_1=-\frac{2\pi T E_{\nu}^2{E_{\bar{\nu}}}^2\left(1-\cos{\theta}\right)^2}
{ \left[\exp (E_{\nu}+E_{\bar{\nu}})/T)-1\right]{\Delta_e}^5}
  \left \{ AT^2\left(\left[G_2\left(y_{\mathrm{max}}\right)-G_2\left(y_{\mathrm{min}}\right)\right] \right.\right. \nonumber \\
&&\ \ \ \ \ \ \ \ \ \ \ \ \ \ \ \ \ \ \ \ \ \ \ \ \ \ \ \ \ \left.\left. +\left[2y_{\mathrm{max}}G_1\left(y_{\mathrm{max}}\right)-2y_{\mathrm{min}}G_1\left(y_{\mathrm{min}}\right)\right]+\left[y_{\mathrm{max}}^2G_0\left(y_{\mathrm{max}}\right)-y_{\mathrm{min}}^2G_0\left(y_{\mathrm{min}}\right)\right]\right)  \right. \nonumber \\
&&\left. +BT\left(\left[G_1\left(y_{\mathrm{max}}\right)-G_1\left(y_{\mathrm{min}}\right)\right]+\left[y_{\mathrm{max}}G_0\left(y_{\mathrm{max}}\right)-y_{\mathrm{min}}G_0\left(y_{\mathrm{min}}\right)\right]\right)+C\left(G_0\left(y_{\mathrm{max}}\right)-G_0\left(y_{\mathrm{min}}\right)\right) \right \}, \nonumber \\
&&
\end{eqnarray}
with $\eta^{\prime}=\left(\mu_e+E_{\nu}+E_{\bar{\nu}}\right)/T$, $\eta=\mu_e/T,y_{\mathrm{max}}=E_{\mathrm{max}}/T$, $y_{\mathrm{min}}=E_{\mathrm{min}}/T$ and 
$G_n\left(y\right)\equiv F_n\left(\eta^{\prime}-y\right)-F_n\left(\eta-y\right)$. 

Similar calculations can be done for the other two integrals to give
\begin{eqnarray}
I_2&=&I_1\left(E_{\bar{\nu}},E_{\nu},\cos{\theta}\right), \\
I_3&=&-\frac{2\pi Tm_e^2 E_{\nu}{E_{\bar{\nu}}}\left(1-\cos{\theta}\right)}
{\left[\exp(E_{\nu}+E_{\bar{\nu}}/T)-1\right] \Delta_e}\left[G_0\left(y_{\mathrm{max}}\right)-G_0\left(y_{\mathrm{min}}\right)\right].
\end{eqnarray}

\section{reaction rate of plasmon decay}

The properties of plasmon are derived from the so-called polarization tensor $\Pi^{\mu\nu}$, which is calculated field-theoretically as
\begin{eqnarray}
i\Pi^{\mu\nu}(K)=\frac{4i}{e^2}\int \frac{d^3k}{2E_e\left(2\pi\right)^3} \frac{(k\cdot K)\left(K^\mu k^\nu+K^\nu k^\mu\right)
- K^2k^\mu k^\nu-(k\cdot K)^2g^{\mu\nu}}{\left(k\cdot K\right)^2}\left(f_{e^-}\left(E_e\right)+f_{e^+}\left(E_e\right)\right), \ \ 
\end{eqnarray}
in which $K=\left(\omega, \mbox{\boldmath$K$}\right)$ and $k=\left(E_e, \mbox{\boldmath$k$}\right)$ are the 4-momenta of plasmon and electron, respectively. 
It is decomposed into the transverse ($\Pi_T$) and longitudinal ($\Pi_L$) components, which are expressed as
\begin{equation}
\Pi^{\mu\nu}(K)=\Pi_T (K) P_T^{\mu\nu}(K) +\Pi_L(K) P_L^{\mu\nu}(K),
\end{equation}
where the projection operators $P_T^{\mu\nu}$ and $P_L^{\mu\nu}$ are given as
\begin{eqnarray}
{P_T}^{\mu\nu} (K) & = & \left \{
\begin{array}{l}
0 \ \ \ \ \ \ \ \ \ \ \ \ \ \ \ \ \ \ (\mu, \nu)=(0, 0)\\
0 \ \ \ \ \ \ \ \ \ \ \ \ \ \ \ \ \ \ (\mu, \nu)=(0, i)\\
\displaystyle{-\delta^{ij}+\frac{K^iK^j}{|\mbox{\boldmath$K$}|^2}}  \ \ \ \ (\mu, \nu)=(i, j)
\end{array}
\right. ,\\
{P_L}^{\mu\nu} (K) & = &  \left \{
\begin{array}{l}
\displaystyle{-\frac{|\mbox{\boldmath$K$}|^2}{K^2}} \ \ \ \ \ \ \ \ \ \ \ \ (\mu, \nu)=(0, 0)\\
\displaystyle{-\frac{\omega K^i}{K^2}} \ \ \ \ \ \ \ \ \ \ \ (\mu, \nu)=(0, i)\\
\displaystyle{-\frac{\left(\omega \right)^2K^iK^j}{K^2|\mbox{\boldmath$K$}|^2}} \ \ \ \ (\mu, \nu)=(i, j)
\end{array}
\right. .
\end{eqnarray}
Using the relations $P_T^{\mu\nu}{P_L}_{\mu\nu}=0$, $P_T^{\mu\nu}{P_T}_{\mu\nu}=2$, $P_L^{\mu\nu}{P_L}_{\mu\nu}=1$ between the projection operators, 
we obtain each component of the polarization tensor as
\begin{eqnarray}
\Pi_T(K) & = & \frac{1}{2}\Pi^{\mu\nu}(K){P_T}_{\mu\nu} (K) \nonumber \\
&=&-\frac{2}{e^2} \int \!\! \frac{d^3k}{2E_e\left(2\pi\right)^3} \frac{-K^2 \left| \mbox{\boldmath$k$} \right|^2 + \displaystyle{\frac{K^2}{\left|\mbox{\boldmath$K$}\right|^2}}
\left(\mbox{\boldmath$K$}\cdot\mbox{\boldmath$k$}\right)^2+2\left(k\cdot K\right)^2}
{\left(k\cdot K\right)^2}\left(f_{e^-}\left(E_e\right)+f_{e^+}\left(E_e\right)\right), \ \ \\
\Pi_L(K) & = & - \frac{K^2}{\left|\mbox{\boldmath$K$}\right|^2}\Pi^{00} \nonumber \\
&=& - \frac{4}{e^2} \frac{K^2} {\left|\mbox{\boldmath$K$}\right|^2} \int \!\! \frac{d^3k}{2E_e\left(2\pi\right)^3} 
\frac{-K^2{E_e}^2+2\left(k\cdot K\right)\left(E_e \, \omega\right) - \left(k\cdot K\right)^2}{\left(k\cdot K\right)^2} \left(f_{e^-}\left(E_e\right)+f_{e^+}\left(E_e\right)\right). \ \ \ \ \ \ \ 
\end{eqnarray}
In the relativistic limit, i.e.,  $E_e \gg \left|\mbox{\boldmath$K$}\right|, \omega$ and $m_e \rightarrow 0$ and $E_e\sim |\mbox{\boldmath$k$}|$, which is well justified
in the present case, the above expressions are reduced to the following:
\begin{eqnarray}
\Pi_T(K) & = & -\frac{3}{2}\frac{\omega^2}{e^2\left|\mbox{\boldmath$K$}\right|^2}{\omega_p}^2
\left [1-\frac{\omega^2-\left|\mbox{\boldmath$K$}\right|^2}{\omega^2}\frac{\omega}{2\left|\mbox{\boldmath$K$}\right|}
\ln{\frac{\omega+\left|\mbox{\boldmath$K$}\right|}{\omega-\left|\mbox{\boldmath$K$}\right|}}\right], \\
\Pi_L(K) & = & -3{\omega_p}^2\frac{\omega^2-\left|\mbox{\boldmath$K$}\right|^2}{e^2\left|\mbox{\boldmath$K$}\right|^2}
\left[\frac{\omega}{2\left|\mbox{\boldmath$K$}\right|}\ln{\frac{\omega+\left|\mbox{\boldmath$K$}\right|}{\omega-\left|\mbox{\boldmath$K$}\right|}}-1\right],
\end{eqnarray}
in which the plasma frequency $\omega_p$ is defined as
\begin{equation}
{\omega_p}^2=\frac{2e^2}{3\pi^2}\left(k_BT\right)^2\left(F_1\left(\eta\right)+F_1\left(-\eta\right)\right).
\end{equation}

The number of the reaction per unit time and volume, $R$, to produce a pair of neutrino and anti-neutrino via the decay of a plasmon is given by the polarization tensor $\Pi^{\mu\nu}$
as follows:
\begin{eqnarray}
R&=&\left(\frac{G}{\sqrt{2}}\right)^2 8\left(q_\mu {q_\nu}^\prime+q_\nu{q_\mu}^\prime-g_{\mu\nu}q\cdot q^\prime\right)2\pi {C_v}^2K^2
\left[\Theta(\omega)\left(1+f_B(\omega)\right)+\Theta(-\omega)f_B(-\omega)\right] \nonumber \\
&&\ \ \ \ \ \ \ \ \ \ \ \ \ \ \ \ \ \ \ \ \ \ \ \ \times\left \{\Pi_L(K) P_L^{\mu\nu}\delta\left(K^2+e^2\Pi_L(K)\right)+\Pi_T(K)P_T^{\mu\nu}\delta\left(K^2+e^2\Pi_T(K)\right) \right \} \nonumber \\
& = & \displaystyle{\left(\frac{G}{\sqrt{2}}\right)^2\frac{8K^2}{\left[1-\exp\left(-\omega / k_BT \right)\right]}\left[\Theta\left(\omega\right)
- \Theta\left(-\omega\right)\right]\left(q_\mu q_\nu^\prime+q_\nu q_\mu^\prime-g_{\mu\nu}q\cdot q^\prime\right)} \nonumber \\
&&\ \ \ \times\left \{ -\frac{3}{2}\frac{\omega^2}{e^2\left|\mbox{\boldmath$K$}\right|^2}{\omega_p}^2
\left [1-\frac{\omega^2-\left|\mbox{\boldmath$K$}\right|^2}{\omega^2}\frac{\omega}{2\left|\mbox{\boldmath$K$}\right|}
\ln{\frac{\omega+\left|\mbox{\boldmath$K$}\right|}{\omega-\left|\mbox{\boldmath$K$}\right|}}\right] P_T^{\mu\nu}\delta\left(K^2+e^2\Pi_T(K)\right)  \right.   \nonumber \\
&&\ \ \ \ \ \ \ \ \ \ \ \ \ \ \ \ \ \ \ \ \ \left. -3{\omega_p}^2\frac{\omega^2-\left|\mbox{\boldmath$K$}\right|^2}{e^2\left|\mbox{\boldmath$K$}\right|^2}
\left[\frac{\omega}{2\left|\mbox{\boldmath$K$}\right|}\ln{\frac{\omega+\left|\mbox{\boldmath$K$}\right|}{\omega-\left|\mbox{\boldmath$K$}\right|}}
-1\right]P_L^{\mu\nu}\delta\left(K^2+e^2\Pi_L(K)\right)\right \}. \ \ \ \ \ \ \ \ \ 
\end{eqnarray}
Note that the dispersion relations for the transverse and longitudinal plasmons are obtained from the $\delta$-functions as $\omega=\omega_T(\mbox{\boldmath$K$})$ and
 $\omega=\omega_L(\mbox{\boldmath$K$})$, which satisfy the following equations:
\begin{eqnarray}
&&K^2+e^2\Pi_T(K)= K^2 - \frac{3}{2}{\omega_p}^2\frac{{\omega_T}^2}{\left|\mbox{\boldmath$K$}\right|^2}
\left [1-\frac{{\omega_T}^2-\left|\mbox{\boldmath$K$}\right|^2}{{\omega_T}^2}\frac{\omega_T}{2\left|\mbox{\boldmath$K$}\right|}
\ln{\frac{\omega_T+\left|\mbox{\boldmath$K$}\right|}{\omega_T-\left|\mbox{\boldmath$K$}\right|}}\right]=0, \\
&&K^2+e^2\Pi_L(K) = K^2 - 3{\omega_p}^2\frac{K^2}{\left|\mbox{\boldmath$K$}\right|^2}
\left[\frac{\omega_L}{2 \left|\mbox{\boldmath$K$}\right|} \ln{\frac{\omega_L+\left|\mbox{\boldmath$K$}\right|}{\omega_L-\left|\mbox{\boldmath$K$}\right|}}-1 \right]=0.
\end{eqnarray}

Employing the conservation law $K = -\left(q+q^\prime\right)$, we finally obtain the reaction rate $R$ as a function of $E_\nu, E_{\bar{\nu}}, \cos{\theta}$ as follows:
\begin{eqnarray}
R & = &\left(\frac{G}{\sqrt{2}}\right)^2\frac{16{C_V}^2}{e^2}\frac{2{E_\nu}^2{E_{\bar{\nu}}}^2
\left(1-\cos{\theta}\right)}{\left[1-\exp\left(E_\nu+E_{\bar{\nu}})/k_BT\right)\right]} \nonumber \\
&&\times \left \{\frac{3{\omega_p}^2}{{\Delta_e}^2} \delta\left(f_L\left(E_\nu, E_{\bar{\nu}}, \cos{\theta}\right)\right)\left[\frac{E_\nu+E_{\bar{\nu}}}{2\Delta_e}\ln{\frac{E_\nu+E_{\bar{\nu}}-\Delta_e}{E_\nu+E_{\bar{\nu}}+\Delta_e}} +1\right]  \right. \nonumber \\
&&\left. \ \ \ \ \times \left[-2\cos{\theta}\left(E_\nu+E_{\bar{\nu}}\right)^2-2E_\nu E_{\bar{\nu}}{\sin{\theta}}^2+\frac{2\left(E_\nu+E_{\bar{\nu}}\right)^2}{{\Delta_e}^2}\left(E_\nu+E_{\bar{\nu}}\cos{\theta}\right)\left(E_{\bar{\nu}}+E_\nu\cos{\theta}\right)\right]  \right. \nonumber \\
&&\left. \ \ -\frac{3{\omega_p}^2\left(E_\nu+E_{\bar{\nu}}\right)^2}{{\Delta_e}^2} \delta\left(f_T\left(E_\nu, E_{\bar{\nu}}, \cos{\theta}\right)\right)\left[1+\frac{E_\nu E_{\bar{\nu}}\left(1-\cos{\theta}\right)}{\left(E_\nu+E_{\bar{\nu}}\right)\Delta_e}\ln{\frac{E_\nu+E_{\bar{\nu}}-\Delta_e}{E_\nu+E_{\bar{\nu}}+\Delta_e}}\right] \right. \nonumber \\
&&\ \ \ \ \left.  \times\left[1-\frac{\left(E_\nu\cos{\theta}+E_{\bar{\nu}}\right)\left(E_{\bar{\nu}}\cos{\theta}+E_\nu\right)}{{\Delta_e}^2}\right]  \right \},
\end{eqnarray}
with $f_L\left(E_\nu, E_{\bar{\nu}}, \cos{\theta}\right)$ and $f_T\left(E_\nu, E_{\bar{\nu}}, \cos{\theta}\right)$ given as
\begin{eqnarray}
&&f_L\left(E_\nu, E_{\bar{\nu}}, \cos{\theta}\right)=2E_\nu E_{\bar{\nu}}\left(1-\cos{\theta}\right) \nonumber \\
&&\ \ \ \ \ \ \ \ \ \ \ \ \ \ \ \ \ \ \ \ \ \ \ +3{\omega_p}^2\frac{2E_\nu E_{\bar{\nu}}\left(1-\cos{\theta}\right)}{{\Delta_e}^2}\left \{\frac{E_\nu+E_{\bar{\nu}}}{2\Delta_\nu}\ln{\frac{E_\nu+E_{\bar{\nu}}-\Delta_e}{E_\nu+E_{\bar{\nu}}+\Delta_e}}+1 \right \}, \\
&&f_T\left(E_\nu, E_{\bar{\nu}}, \cos{\theta}\right)=2E_\nu E_{\bar{\nu}}\left(1-\cos{\theta}\right) \nonumber \\
&&\ \ \ \ \ \ \ \ \ \ \ \ \ \ \ \ \ \ \ \ \ \ \ -\frac{3}{2}{\omega_p}^2\frac{\left(E_\nu+E_{\bar{\nu}}\right)^2}{{\Delta_e}^2}\left[1+\frac{E_\nu E_{\bar{\nu}}\left(1-\cos{\theta}\right)}{\left(E_\nu+E_{\bar{\nu}}\right)\Delta_e}\ln{\frac{E_\nu+E_{\bar{\nu}}-\Delta_e}{E_\nu+E_{\bar{\nu}}+\Delta_e}}\right]. \ \ \ 
\end{eqnarray}

\acknowledgments{
This work is partly supported by Grant-in-Aid for Scientific Research from the Ministry of Education, Culture, Sports, Science and Technology of Japan 
(Nos. 24244036, 24103006, 26104007, 26400220), and HPCI Strategic Program of Japanese MEXT. K.T. is supported by Research Fellowships of 
Japan Society for the Promotion of Science (JSPS) for Young Scientists. }

\clearpage

\clearpage
\begin{table}[H]
\begin{center}
\large
\caption{The detector parameters assumed in this paper.\tablenotemark{a}\label{detector}}
\begin{tabular}{cccccc}
\tableline\tableline
Detector &$\ \ \ \ \ \ \ $ Mass &$\ \ \ \ \ \ \ $ Target number &$\ \ \ \ \ \ \ $ Energy threshold  \\
&$\ \ \ \ \ \ \ $[kt]&$\ \ \ \ \ \ \ $N&$\ \ \ \ \ \ \ $[MeV] \\
\tableline
Super-K & $\ \ \ \ \ \ \ $32 &$\ \ \ \ \ \ \ $2.14$\times 10^{33}$ &$\ \ \ \ \ \ \ $5.3 \\
KamLAND&$\ \ \ \ \ \ \ $1 &$\ \ \ \ \ \ \ $8.47$\times 10^{31}$ &$\ \ \ \ \ \ $ 1.8\\
Hyper-K &$\ \ \ \ \ \ \ $540  &$\ \ \ \ \ \ \ $3.61$\times 10^{34}$ &$\ \ \ \ \ \ \ $8.3 & \\
JUNO&$\ \ \ \ \ \ \ $20&$\ \ \ \ \ \ \ $1.69$\times 10^{33}$&$\ \ \ \ \ \ \ $1.8\\
\tableline
\end{tabular}
\tablenotetext{a}{The numbers given here are not very precise and just  meant for rough estimate. JUNO is assumed to be a scale-up of KamLAND by a factor of 20. We also assume that the energy threshold of Hyper-Kamiokande will be somewhat higher than that of Super-Kamiokande.}
\tablerefs{
(1) \citealt{sk}; (2) \citealt{kam}; (3) \citealt{hk}; (4) \citealt{juno14}}
\end{center}
\end{table}

\begin{table}[H]
\begin{center}
\large
\caption{The expected numbers of detection events for different detectors.\tablenotemark{b} \label{eventrate}}
\begin{tabular}{ccccccccccc}
\tableline\tableline
detector & \multicolumn{2}{c}{8.4 $\ {\rm M_\odot}$} & \multicolumn{2}{c}{12 $\ {\rm M_\odot}$} & \multicolumn{2}{c}{15 $\ {\rm M_\odot}$}  \\
&normal&inverted&normal&inverted&normal&inverted \\
\tableline
Super-K & 2.47 $\times 10^{-2}$ &9.68$\times 10^{-3}$ & 21 & 7 & 61 & 21\\
KamLAND&1.06  $\times 10^{-3}$& 1.50$\times 10^{-3}$ & 31 & 9 & 43 & 13 \\
Hyper-K &0.30 & 0.13 & 9 & 3 & 77 & 28 \\
JUNO&2.12 $\times 10^{-2}$ & 8.03$\times 10^{-3}$ & 618 & 189 & 864 & 266\\
\tableline
\end{tabular}
\tablenotetext{b}{The source is assumed to be located at $200\ {\rm pc}$ from the earth. Both the normal and inverted mass hierarchies are considered in the adiabatic oscillation limit.}
\end{center}
\end{table}

\end{document}